\documentclass[twocolumn,showpacs,aps,prd,nofootinbib]{revtex4-1}

\usepackage{longtable}
\usepackage{tabularx}

\usepackage{graphics}
\usepackage{amssymb}
\usepackage{amsthm}
\usepackage[mathlines]{lineno}
\usepackage{graphicx}
\usepackage{units}
\usepackage{url}
\usepackage{amsmath}
\usepackage{amsfonts}
\usepackage{bm}
\usepackage{textcomp}
\usepackage{subfigure}
\usepackage{verbatim}
\usepackage{booktabs}
\usepackage{multirow}
\usepackage{xspace}
\usepackage[colorlinks=true,linkcolor=blue]{hyperref}

\newcommand{\BR}{{\cal B}}

\newcommand{\ee}{e^{+}e^{-}}
\newcommand{\gam}{\gamma}
\newcommand{\gami}{\gamma_{\rm ISR}}

\newcommand{\GeV}{\,\mathrm{GeV}}
\newcommand{\MeV}{\,\mathrm{MeV}}
\newcommand{\keV}{\,\mathrm{keV}}
\newcommand{\mev}{\,\unit{MeV}/c^2}

\newcommand{\gev}{\,\unit{GeV}/c^2}

\newcommand{\gevc}{\,\mathrm{GeV}/c}

\newcommand{\ipb}{\,\mathrm{pb}^{-1}}
\newcommand{\ifb}{\,\mathrm{fb}^{-1}}

\newcommand{\dit}{\tau^{+}\tau^{-}}
\newcommand{\ddb}{D\bar{D}}

\newcommand{\kk}{K^{+}K^{-}}

\newcommand{\mumu}{\mu^{+}\mu^{-}}
\newcommand{\pipi}{\pi^{+}\pi^{-}}
\newcommand{\omegap}{\omega(1420)}
\newcommand{\omegapp}{\omega(1650)}
\newcommand{\fCtC}{5C\, (2C)}
\newcommand{\pip}{\pi^{+}}
\newcommand{\pim}{\pi^{-}}
\newcommand{\piz}{\pi^{0}}
\newcommand{\ppb}{p\bar{p}}

\newcommand{\jpsi}{J/\psi}
\newcommand{\psip}{\psi^{\prime}}
\newcommand{\psipp}{\psi(3770)}
\newcommand{\qqb}{q\bar{q}}

\newcommand{\zmev}{\mathrm{MeV}/c^{2}}

\newcommand{\Kbar    }{\kern 0.2em\overline{\kern -0.2em K}{}\xspace}
\newcommand{\Dbar    }{\kern 0.2em\overline{\kern -0.2em D}{}\xspace}

\newcommand{\Kz      }{\ensuremath{K^0}\xspace}
\newcommand{\Kzb     }{\ensuremath{\Kbar^0}\xspace}
\newcommand{\KzKzb   }{\ensuremath{\Kz \kern -0.16em \Kzb}\xspace}
\newcommand{\Kp      }{\ensuremath{K^+}\xspace}
\newcommand{\Km      }{\ensuremath{K^-}\xspace}

\newcommand{\KpKm    }{\ensuremath{\Kp \kern -0.16em \Km}\xspace}

\newcommand{\tev}{\ensuremath{\mathrm{\,Te\kern -0.1em V}}\xspace}
\newcommand{\kev}{\ensuremath{\mathrm{\,ke\kern -0.1em V}}\xspace}
\newcommand{\ev}{\ensuremath{\mathrm{\,e\kern -0.1em V}}\xspace}

\newcommand{\bei}{\begin{itemize}}
\newcommand{\eei}{\end{itemize}}
\newcommand{\ben}{\begin{enumerate}}
\newcommand{\een}{\end{enumerate}}

\renewcommand{\arraystretch}{1.25}

\def\PL{{Phys. Lett.}\xspace}
\def\PR{{Phys. Rev.}\xspace}

\def\PRL{{Phys. Rev. Lett.}\xspace}
\def\NIM{{Nucl. Inst. Meth.}\xspace}
\def\ZP{{Z. Phys.}\xspace}
\def\EPJ{{Eur. Phys. J.}\xspace}
\def\CPC{{Comput. Phys. Commun.}\xspace}

\def\ChP{Chin. Phys.}
\def\etal{\emph{et al.}~}

\def\Journal#1#2#3#4{{#1} {\bf #2}, #3 (#4)}

\usepackage{relsize}
\def\babar{\mbox{\slshape B\kern-0.1em{\smaller A}\kern-0.1em B\kern-0.1em{\smaller A\kern-0.2em R}}\,}

\begin{document}
\interfootnotelinepenalty=10000

\title{Measurement of the  $\ee\to\pipi\piz$ cross section from 0.7 GeV to 3.0 GeV\\ via initial-state radiation}

\author{
  \begin{small}
\begin{center}
	M.~Ablikim$^{1}$, M.~N.~Achasov$^{10,d}$, P.~Adlarson$^{63}$, S. ~Ahmed$^{15}$, M.~Albrecht$^{4}$, M.~Alekseev$^{62A,62C}$, A.~Amoroso$^{62A,62C}$, F.~F.~An$^{1}$, Q.~An$^{59,47}$, Y.~Bai$^{46}$
% ...
\\
\vspace{0.2cm}
(BESIII Collaboration)\\
\vspace{0.2cm} {\it
$^{1}$ Institute of High Energy Physics, Beijing 100049, People's Republic of China\\
% ...
$^{4}$ Bochum Ruhr-University, D-44780 Bochum, Germany\\
% ...
$^{10}$ G.I. Budker Institute of Nuclear Physics SB RAS (BINP), Novosibirsk 630090, Russia\\
% ...
$^{15}$ Helmholtz Institute Mainz, Johann-Joachim-Becher-Weg 45, D-55099 Mainz, Germany\\
% ...
$^{46}$ Southeast University, Nanjing 211100, People's Republic of China\\
$^{47}$ State Key Laboratory of Particle Detection and Electronics, Beijing 100049, Hefei 230026, People's Republic of China\\
% ...
$^{59}$ University of Science and Technology of China, Hefei 230026, People's Republic of China\\
% ...
$^{62}$ (A)University of Turin, I-10125, Turin, Italy; (B)University of Eastern Piedmont, I-15121, Alessandria, Italy; (C)INFN, I-10125, Turin, Italy\\
$^{63}$ Uppsala University, Box 516, SE-75120 Uppsala, Sweden\\
% ...
$^{d}$ Also at the Novosibirsk State University, Novosibirsk, 630090, Russia\\
% ...
}\end{center}
\end{small}
}
\noaffiliation{}

\begin{linenomath}
\begin{abstract}
The cross section of the process $\ee\to\pipi\piz$ is measured  with a precision of 1.6\% to 25\% in the energy range  between $0.7$ and 3.0 GeV using the Initial State Radiation method.
A data set with an integrated luminosity of  $2.93\ifb$ taken at the center-of-mass energy of $\sqrt{s}=3.773\GeV$ with the BESIII detector at the BEPCII collider is used. 
The product branching fractions for 
$\omega$, $\phi$, $\omegap$, and $\omegapp$ are measured to be 
$\BR(\omega\to\ee)\times\BR(\omega\to\pipi\piz)=(6.94\pm0.08\pm0.16)\times10^{-5}$, 
$\BR(\phi\to\ee)\times\BR(\phi\to\pipi\piz)=(4.20\pm0.08\pm0.19)\times10^{-5}$, 
$\BR(\omegap\to\ee)\times\BR(\omegap\to\pipi\piz)=(0.84\pm0.09\pm0.09)\times10^{-6}$, 
and $\BR(\omegapp\to\ee)\times\BR(\omegapp\to\pipi\piz)=(1.14\pm0.15\pm0.15)\times10^{-6}$, respectively. The branching 
fraction $\mathcal{B}(J/\psi\to\pipi\piz)$ is measured to be 
  $(2.188\pm0.024\pm0.024\pm0.040(\Gamma_{ee}^{\jpsi}))\%$, where $\Gamma_{ee}^{\jpsi}$ is the dileptonic width of $\jpsi$. 
The first errors are of statistical, the second and third ones of systematic nature.
\end{abstract}
\end{linenomath}



\maketitle

\section{Introduction}\label{s_int}
The Standard Model~(SM) of particle physics has so far successfully withstood most of the challenges it has been confronted with. 
However, there are still deficiencies in the SM, and the search for
physics beyond the SM (`New Physics') is a major effort nowadays.
A hint for New Physics might come from the anomalous magnetic moment of the 
muon, $a_\mu \equiv \frac{1}{2}(g_\mu-2)$. A discrepancy of $(27.06\pm7.26)\times10^{-10}$, corresponding to 3.7 standard 
deviations, between the direct measurement~\cite{c_bnl} and the SM prediction~\cite{c_thomas_2018} has been found. New experiments 
are expected to reach a precision of 0.14 ppm at Fermilab~\cite{c_g-2} and at J-PARC~\cite{c_jparc}. At the same time, the theoretical accuracy is completely 
limited by the hadronic contributions to $a_\mu$. The largest of these contributions is due to the hadronic Vacuum 
Polarization~(hVP). 

Although perturbative Quantum Chromodynamics~(QCD) theory fails in the energy regime relevant for the calculation of $a_\mu$,
 it is possible to relate the leading order hVP contribution $a_\mu^{\rm hVP}$ to hadronic 
production rates in $\ee$ annihilations via the dispersion relation
\begin{linenomath}
 \begin{equation}
  \label{eq_dis}
   a_\mu^{\rm hVP}=\frac{\alpha^2}{3\pi^2}\int_{4m_{\pi}^{2}}^{\infty} R_{\rm had}(s)\times K(s){\rm d}s,
 \end{equation}
\end{linenomath}
where $\alpha$ is the fine-structure constant, $m_\mu$ is the muon mass, $s$ is the center-of-mass energy squared, $R_{\rm had}(s)$ represents the cross section ratio of $\ee\to{\rm hadrons}$ to $\ee\to\mumu$, and $K(s)$ is a 
kernel function~\cite{c_3pi_Ks}, which is monotonously decreasing with increasing  
$s$. The energy dependence of both the hadronic cross section and the kernel function imply that $a_{\mu}^{\rm hVP}$ is more sensitive to  the $\ee\to{\rm hadrons}$ cross section in the 
low energy region. 

In order to improve the SM prediction for $a_\mu$, precision measurements of
hadronic cross sections at $\ee$ colliders are needed as input to the dispersive calculations.
This can be accomplished by a direct, precise measurement of the cross section of 
$\ee \to \pipi\piz$ in the energy region from $0.7$ to $3.0\GeV$, and this is the 
main topic for the analysis presented in this report.
In addition to the evaluation of the hadronic 
contributions to $a_{\mu}$, the data allow also the study of the decays of the light vector mesons, $\rho$, $\omega$, $\phi$, and 
excited states.

The hadronic cross section in $\ee$ annihilations at $\sqrt{s}<1.05\GeV$ can be well described  by the Vector Meson 
Dominance~(VMD) model. The ground state vector mesons $\rho$, $\omega$, and $\phi$ are well understood, and their 
properties, such as the mass, width, and main decay modes, are precisely measured~\cite{c_pdg2014}. However, the 
conventional VMD model, taking only into account the ground states, is not sufficient to describe $\ee$ data above 
$\phi$ mass region. Contributions from radial excitations of  possible vector mesons with quantum 
numbers $I^G(J^{PC}) = 1^+(1^{--})$, $0^-(1^{--})$ 
and masses around $\sqrt{s}=1.3\GeV$ and $1.6\GeV$ need to be taken into account. 

At low energies, the $\pipi\piz$ spectrum is dominated by the $\omega$
and $\phi$ mesons. The mass region below 1.4~$\GeV$ has been studied 
by SND~\cite{c_3pi_snd} and \mbox{CMD-2}~\cite{c_3pi_cmd2} with high statistics. At individual 
scan points their results show some deviations, however. Above the peak of the $\phi$ resonance, the \babar  experiment has 
measured the $\ee\to\pipi\piz$~\cite{c_g3pi_babar} cross section up to $\sqrt{s}=3.0\GeV$, exploiting the Initial State 
Radiation (ISR) method. The result agrees with the SND measurement~\cite{c_3pi_snd} for $\sqrt{s}<1.4\GeV$ and 
significantly exceeds the cross section measured by DM2~\cite{c_3pi_dm2} in the region $1.4<\sqrt{s}<2.2\GeV$.

The emission of ISR photons with energies $E_\gamma$ allows the production of the hadronic states at an ``effective'' 
center-of-mass energy $\sqrt{s'}$, well below the actual $\sqrt{s}$ of the $\ee$ collision. Excluding next-to-leading order and final state radiation effects, the actual and the effective 
center-of-mass energies are related according to
$s'=s-2\sqrt{s}E_{\gamma}$. The large data samples 
acquired at recent $\ee$ colliders permit measuring the hadronic cross section in a wide range by exploiting the ISR technique, 
which is complementary to studies at energy scan experiments. The ISR method, originally used both at the KLOE~\cite{c_kloe} experiment 
at DA$\Phi$NE and at the \babar~\cite{c_babar} experiment at the PEP-II B-factory, has been applied to measure numerous 
channels of the cross section $\ee\to{\rm hadrons}$. This allows for a consistent measurement of the full energy range 
with the same accelerator and detector conditions. Different reconstruction techniques are applied according to the 
angular distribution of the ISR photons. Only a small fraction of the photons is emitted with large polar angles and 
can be reconstructed in the detector. The exclusive reconstruction of such events is referred to as ``tagged ISR method'' 
and makes measuring the cross section down to the hadronic mass threshold possible. 
A tagged ISR measurement of the channel $\ee\to\pipi$ has been performed by BESIII recently~\cite{c_kloss}. Most of the ISR photons are emitted 
along the beam direction, which is beyond the acceptance of currently existing detection systems at $\ee$ machines. 
However, it is still possible to reconstruct these events based on energy and momentum conservation, which is referred
to as ``untagged ISR method''. In this work, candidate events are selected both according to the tagged and untagged ISR selections.

\section{The BESIII detector and data samples}\label{s_detector}
The BESIII detector is a magnetic
spectrometer~\cite{Ablikim:2009aa} located at the Beijing Electron
Positron Collider (BEPCII)~\cite{Yu:IPAC2016-TUYA01}. The
cylindrical core of the BESIII detector consists of a helium-based
 multilayer drift chamber (MDC), a plastic scintillator time-of-flight
system (TOF), and a CsI(Tl) electromagnetic calorimeter (EMC),
which are all enclosed in a superconducting solenoidal magnet
providing a 1.0~T 
magnetic field. The solenoid is supported by an
octagonal flux-return yoke with resistive plate counter muon
identifier modules interleaved with steel. The acceptance of
charged particles and photons is 93\% over the $4\pi$ solid angle. The
charged-particle momentum resolution at $1~{\rm GeV}/c$ is
$0.5\%$, and the $dE/dx$ resolution is $6\%$ for the electrons
from Bhabha scattering. The EMC measures photon energies with a
resolution of $2.5\%$ ($5\%$) at $1$~GeV in the barrel (end cap)
region. The time resolution of the TOF barrel part is 68~ps, while
that of the end cap part is 110~ps. 

Two $\ee$ collision data samples were collected with the BESIII detector at $\sqrt{s} = 3.773\GeV$ in 2010 and 2011 and
they are referred to as ``Data I'' and ``Data II''. The corresponding integrated luminosities are determined to be $927.67\pm0.10\pm9.28\ipb$ and $1989.27\pm0.15\pm19.89\ipb$, respectively, by using large-angle Bhabha scattering events ~\cite{c_lum}.

The geometry and response of the BESIII detector are modeled in a {\sc geant{\small4}}-based~\cite{c_geant4} Monte Carlo 
(MC) simulation software. It is used to produce signal and background MC samples for the determination of the mass 
resolution and detection efficiency, as well as for the study of background contributions. The signal MC of 
$\ee\to\gami\pipi\piz$ is modeled in VMD and generated with the {\sc 
phokhara}~\cite{c_phokhara} event generator. The contribution from the $\jpsi$ resonance is not included in this 
generator. Thus, a MC sample of $\ee\to\gami\jpsi\to\gami\pipi\piz$ is generated with the {\sc kkmc}~\cite{c_kkmc} 
and {\sc EvtGen}~\cite{c_evtgen} event generators to study the $\jpsi$ signal, where the {$\rho\pi$} 
model~\cite{c_besevtgen} is used for the decay of $\jpsi\to\pipi\piz$.

The dominating background processes $\ee\to\pipi\piz\piz$ and $\ee\to\gami\pipi\piz\piz$ are generated 
exclusively with {\sc phokhara}~\cite{c_phokhara}. Further, generic MC samples including $\ee\to\qqb$, 
$\ee\to\gami\jpsi$, $\ee\to\gami\psip$, $\psipp\to\ddb$, and $\ee\to\dit$ are generated with {\sc kkmc} 
and {\sc EvtGen}. The subsequent decays of the resonances are processed according to the branching 
fractions provided by the Particle Data Group~(PDG)~\cite{c_pdg2014}. Remaining unmeasured decay modes from charmonium states are generated 
with {\sc lundcharm}~\cite{c_lundc}. The beam energy spread is considered in all MC simulations. Final-State 
Radiation (FSR) effects are simulated with {\sc photos}~\cite{c_photos}.

\section{Event selection}\label{s_sel}
Event candidates of the final state of interest, denoted hereafter as ISR 3$\pi$, contain two charged tracks and three 
photons. For the event selection, exactly two charged tracks are
required to pass the $\ee$ interaction point within $1\,\mathrm{cm}$
in the transverse and within $10\,\mathrm{cm}$ in the longitudinal
direction relative to the beam axis. Each track is required to be reconstructed within 
$|\cos\theta|<0.93$, where $\theta$ is the polar angle with respect to the $e^{+}$ beam direction. The two tracks 
in an event candidate must be oppositely charged. Each track is assumed to be a pion. In order to reduce background 
contributions from Bhabha scattering, both of the tracks are required to satisfy $E/p<0.8$, where $E$ is the 
energy deposited in the EMC and $p$ is the magnitude of the  momentum measured in the MDC.

Photon candidates are selected from showers in the EMC that are not associated with charged tracks. Good photon 
candidates reconstructed from the barrel part of the EMC must have a polar angle within $|\cos\theta|< 0.8$ and  a 
minimum deposited energy of $25\MeV$. To be reconstructed from the end cap, the photon candidates must  have a polar angle within
$0.86<|\cos\theta|<0.92$ and a minimum energy deposit of $50\MeV$. Timing information in the EMC is used 
to suppress electronic noise and energy deposits unrelated to the event. At least two photons matching these criteria 
are required for an analysis using the untagged ISR method, and at least three are required for the tagged ISR method. 
From all possible combinations of two photons, those  which fulfill $\chi^2_{1C}<10$ 
in a fit of the photon candidates constrained to the nominal $\piz$ mass ($1C$) are accepted as $\piz$ candidates.

The criteria of the event selection are optimized for the three-pion invariant mass region $1.05\leq M_{\pipi\piz}\leq2.00\gev$, where the 
discrepancy between \babar and DM2 is most prominent. The optimization
is performed with a figure of merit defined as $\frac{S}{\sqrt{S+B}}$, where $S$ refers to the number of signal events, and $B$ refers to 
the number of events from all background contributions. 
The signal and background MC samples are normalized to the 
data by the integrated luminosity.

A kinematic fit is performed to reject background contributions and to improve the mass resolution. The four-momenta of 
the charged tracks and the photons are constrained to the initial $\ee$ four-momenta. The two photons, which are 
assigned to the $\piz$ candidate, are constrained to the nominal $\piz$ mass. If there are additional photons or 
$\piz$ candidates in a single event, the combination with the minimum $\chi^2$ from the kinematic fit is selected. For 
the tagged ISR method, the constraints add up to $5C$ (four-momenta and $\piz$ mass),
and $\chi^2_{5C}<60$ is required. 
In the untagged ISR method, the radiative photon is considered as an unmeasured massless particle in the kinematic fit. 
Due to the missing momentum information the number of constraints is reduced to a $2C$ 
(missing photon mass and $\piz$ mass) kinematic fit, where 
$\chi^2_{2C}<25$ must be satisfied.

Further reduction of background in the tagged ISR method is achieved by considering the invariant mass distribution of 
the ISR photon and any other good photon in the event
$M_{\gami\gamma}$. Asymmetric $\piz$ decays can lead to misidentified
ISR photons; these wrongly assigned ISR photons result in a peak in
the $M_{\gami\gam}$ distribution and are removed by a $\piz$ veto,
requiring $M_{\gami\gamma}>0.17\gev$.  In the untagged ISR 
method, additional background events are suppressed efficiently with a strict requirement on the scattering angle of the ISR 
photon determined by the kinematic fit, $i.e$  $|\cos\theta_{\gami}|>0.9984$.

Due to the less constrained kinematic fit in the untagged ISR method, a non-negligible contribution of beam-induced background 
 is found.  An additional requirement is therefore applied to 
the untagged ISR method, where the two charged tracks are fitted to a common vertex. The vertex must have a distance to 
the nominal interaction point of less than $2.5\,\textrm{cm}$ in the plane perpendicular to the beams. Background 
contributions attributed to interactions of the beam halo with the
beam pipe are efficiently removed. 
This additional vertex requirement is not requested in the tagged ISR method because such background is rejected with the four-momentum constraint.

\begin{figure*}
	\begin{center}
	\subfigure[Tagged efficiency curve as a function of the true
        mass of $\pipi\piz$ obtained from the generator.]{\label{f_eff:a}
	 \includegraphics[width=0.45\textwidth]{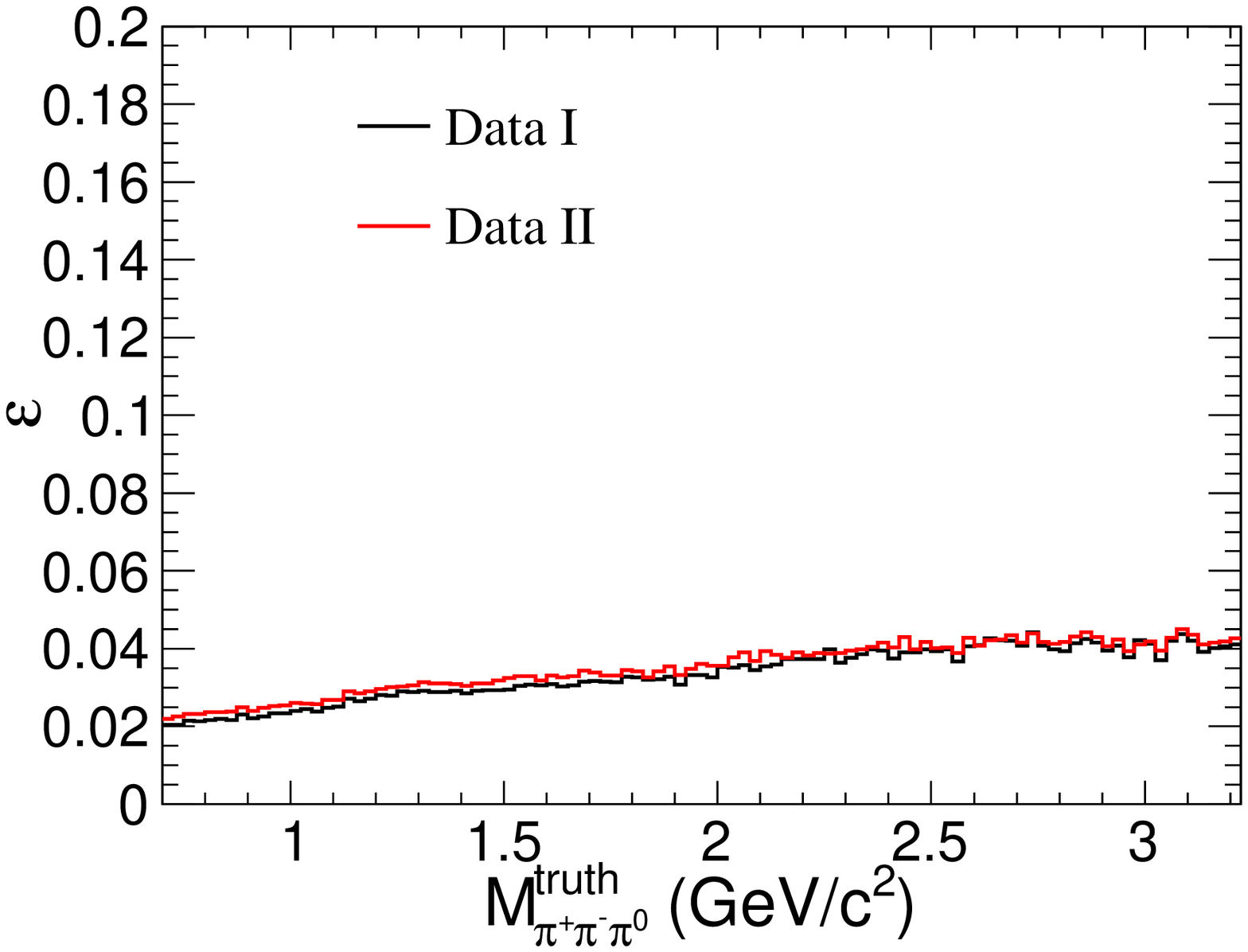}}%
	 \subfigure[Untagged efficiency curve.]{\label{f_eff:b}
	 \includegraphics[width=0.45\textwidth]{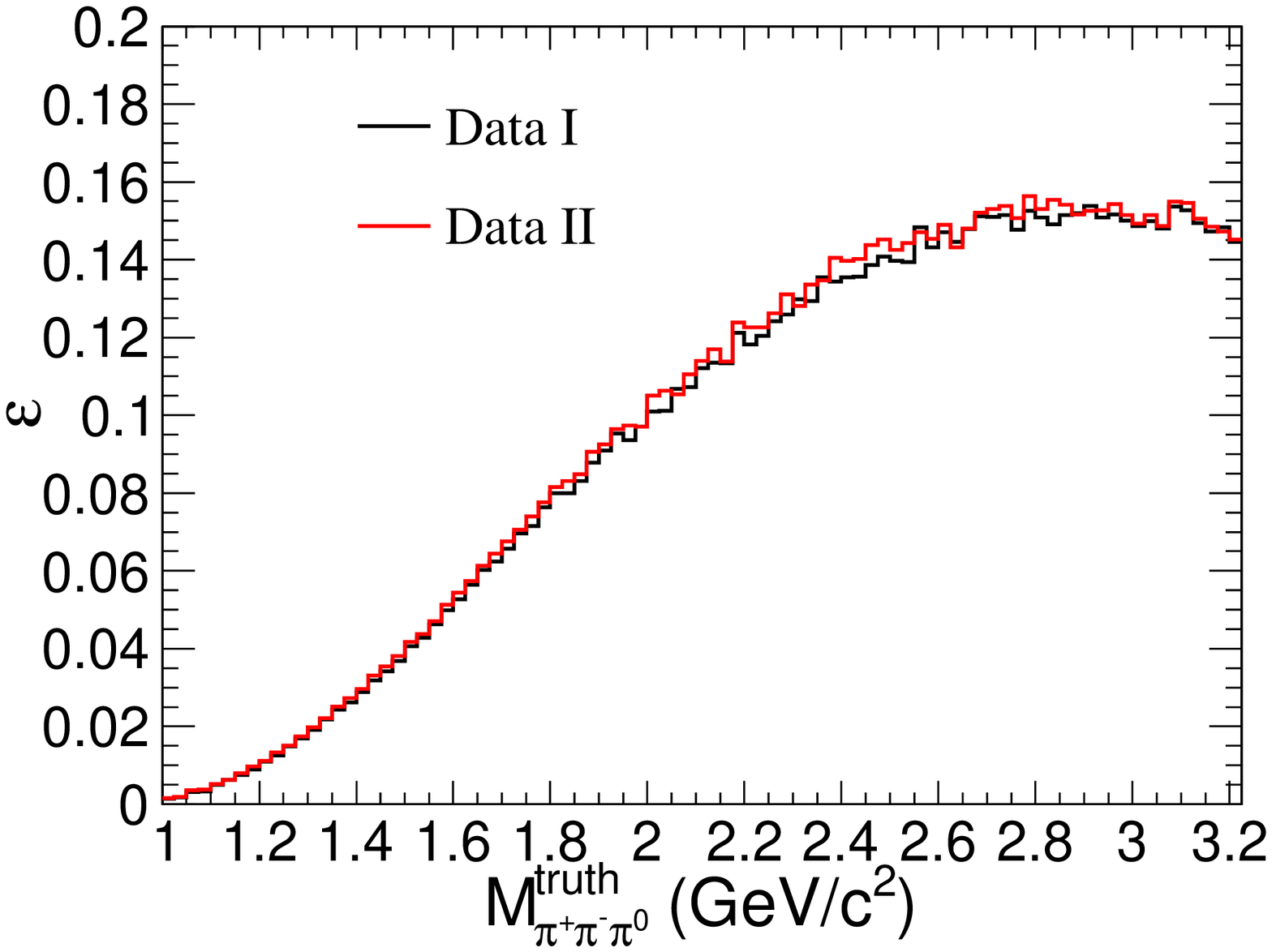}}\\%
		\subfigure[Tagged mass spectrum after 5C kinematic fit.]{\label{f_eff:c}
	  \includegraphics[width=0.45\textwidth]{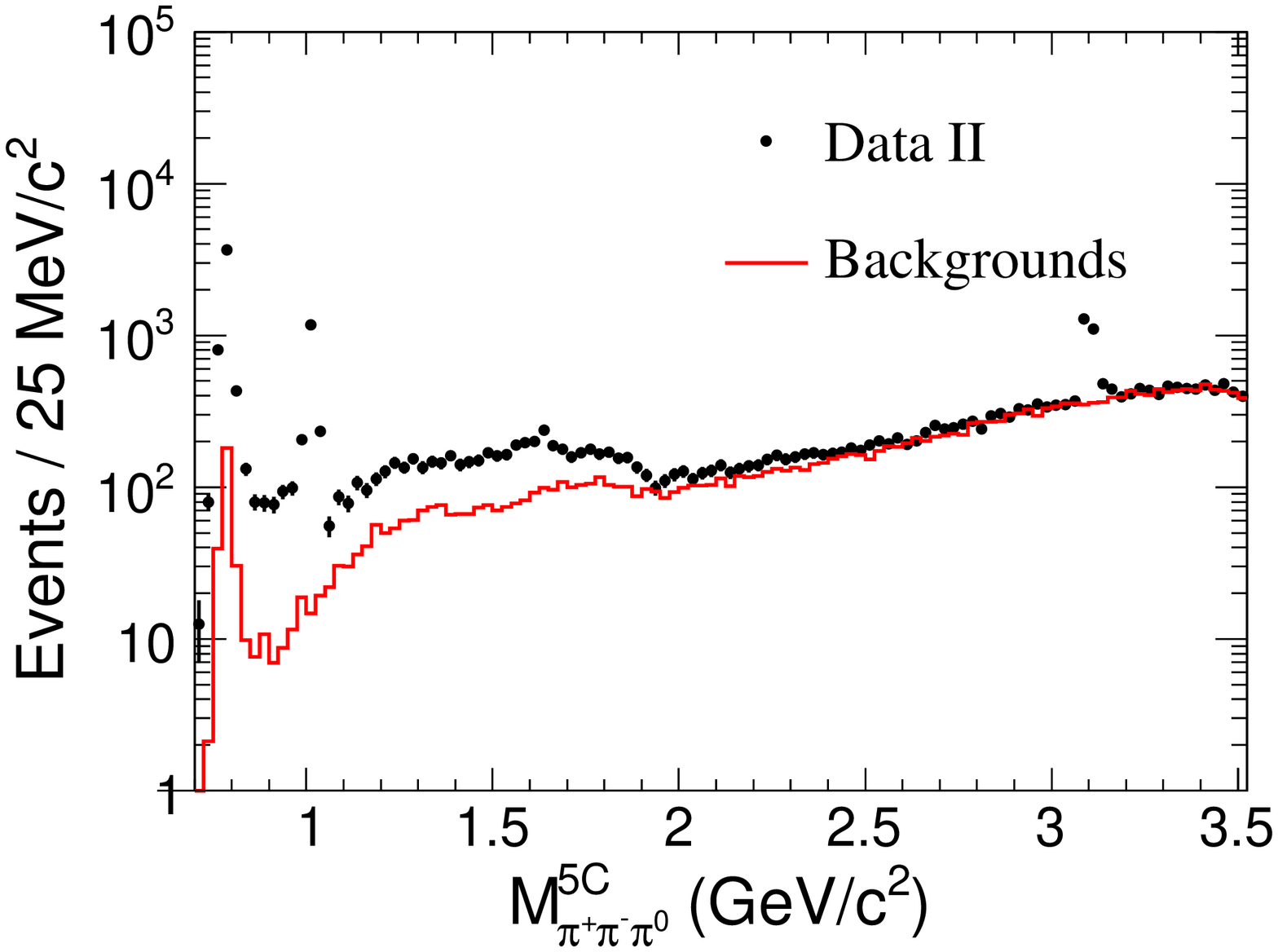}}%
		\subfigure[Untagged mass spectrum after 2C kimatic fit.]{\label{f_eff:d}
	\includegraphics[width=0.45\textwidth]{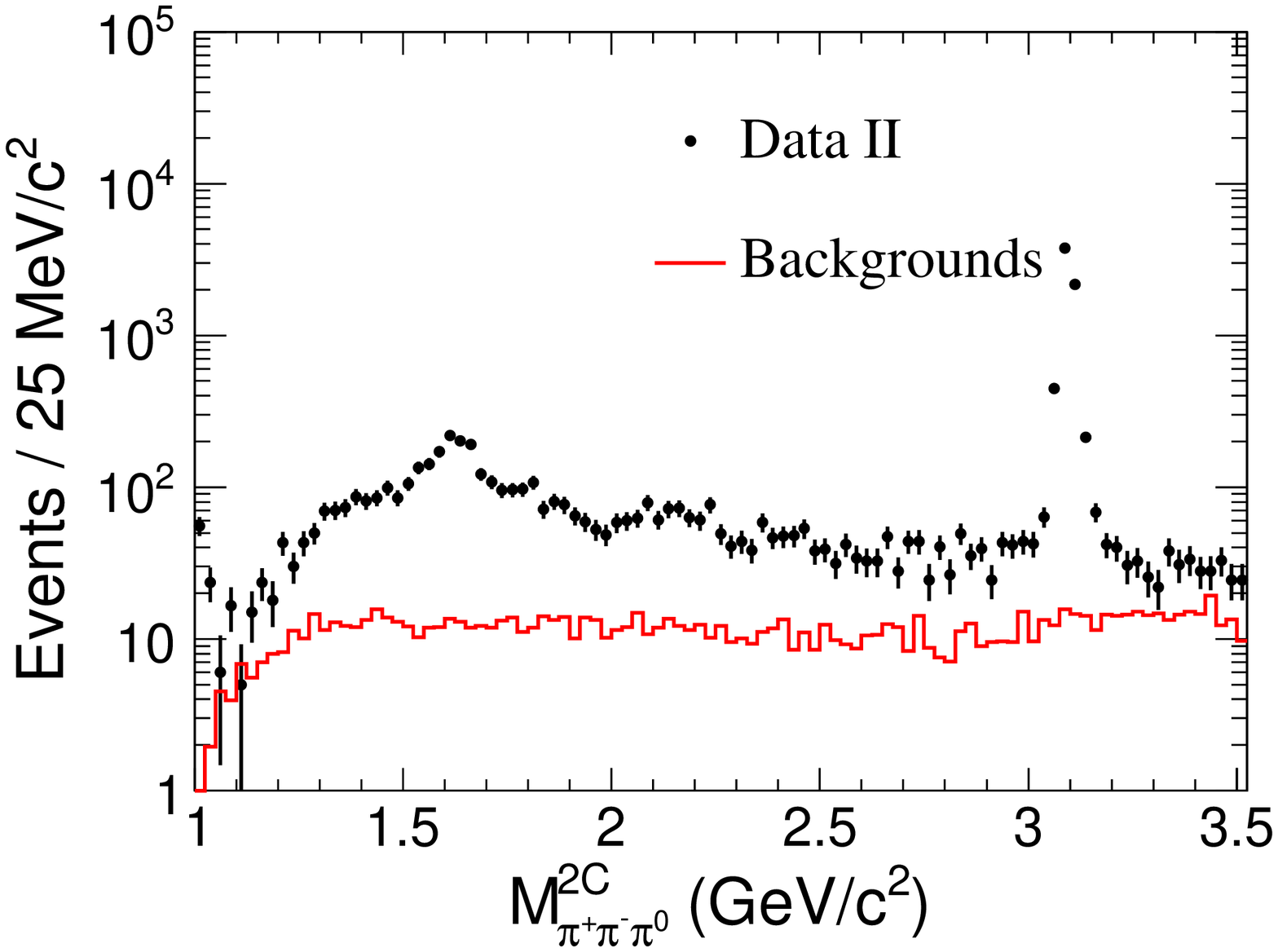}}\\%
		\caption{\label{f_eff} Efficiency curves obtained from
                  simulation (\emph{top row}) and
     mass spectra after the kinematic fit (\emph{bottom row}) for data and background (mass spectra for Data I is similar).
    The left panels are for the tagged ISR selection and the right ones for the untagged selection.
    The main background is estimated using a data-driven method described in Sec.~\ref{s_bg}.
    }

	\end{center}
\end{figure*}

The mass-dependent efficiency is obtained from MC simulations.
It varies from 2\% to 4\% for 
the tagged ISR method in the mass region $0.7\leq M_{\pipi\piz} \leq 3\gev$, and from 0.3\% to 15\% for the untagged ISR 
method for masses between $1.05\gev$ and $3\gev$, as shown in the top row of
Fig.~\ref{f_eff}. The distributions are quite similar for the two data sets.

\section{Background subtraction}\label{s_bg}                                                        
The background contamination of the selected events differs significantly between the tagged and the untagged ISR 
methods. The $3\pi$ mass spectrum obtained from the tagged analysis has a relatively small background contamination at 
masses below $1.05\gev$. At larger masses, the spectrum is dominated by 
background.
The untagged method, 
in comparison, suffers from very low efficiency in the low mass region, but it provides much higher efficiency 
with much less background contamination in the high mass region. 
This can be seen from the red lines in Fig.~\ref{f_eff:c}, \ref{f_eff:d} and 
lines in Fig.~\ref{f_eff:a}, \ref{f_eff:b} which show the background levels and the efficiencies for the tagged and untagged methods, respectively.

Using the $\piz$ side bands is an efficient way to estimate the events with wrongly reconstructed $\piz$ candidates. The $\piz$ 
candidate is constrained to the nominal $\piz$ mass for the signal mass region, and to $0.10\gev$ and $0.17\gev$ for 
the side bands. The $\chi^2_{1C}$ from the $\piz$ mass constraint is required to be  $\chi^2_{1C}<10$ for both signal 
and side bands regions. Contributions from wrongly reconstructed $\piz$ in the signal region are subtracted using 
the averaged number of events in the side bands.

As mentioned earlier, the dominant backgrounds  are due to $\ee\to\pipi\piz\piz$ and $\ee\to\gami\pipi\piz\piz$ (referred to as  
$4\pi$ and ISR 4$\pi$ hereafter). The 4$\pi$ events with a lost soft photon  mainly contaminate the tagged ISR method, 
while the ISR 4$\pi$ events are the dominant background for the untagged ISR method. Other backgrounds contribute to less 
than 1.3\% and are estimated using the generic MC samples.
The dominant backgrounds are estimated using a data-driven approach.

First, a clean 4$\pi$ (ISR 4$\pi$) sample is selected from data. The same selections of 
track and photon candidates as described in Sec.~\ref{s_sel} are used, and at least
four (five) good photons are required for the 4$\pi$ (ISR $4\pi$) sample.
At least two $\piz$ candidates are selected from four good photons by a 
kinematic fit with $\chi^2_{1C}<10$ ($\piz$ mass constraint).
If there are more than two good $\piz$ candidates, the combination with the 
$\pi^+\pi^-$ candidates, which yields the minimum $\chi^2_{6C(3C)}$
(see below) is selected. 
The 4$\pi$ events are selected by constraining the two charged pions 
and two good $\piz$ candidates to a $6C$ (four-momenta conservation plus 
two $\piz$ mass constraints) kinematic fit. A requirement of $\chi^{2}_{6C}<50$ is 
imposed to reject  background contributions to the 4$\pi$ sample.
In order to select ISR $4\pi$ events, the two charged pions, four photons
from two $\piz$ candidates, and one missing 
ISR photon are subjected to a $3C$ kinematic fit. The number of constraints
is reduced by the unknown four-momentum of 
the ISR photon. The $\chi^2_{3C}$ value of the fit is required to be less 
than 20. The direction of the ISR photon
predicted by the kinematic fit is required to satisfy $|\cos\theta|>0.999$.
The reconstructed $4\pi$ mass is required to be less than $3.65\gev$
to reduce contribution of 4$\pi$ events in the ISR 4$\pi$ sample.

Then, a sample of MC simulated $4\pi$ (ISR $4\pi$) events is produced with the {\sc
phokhara}~\cite{c_phokhara} generator. It is used to determine the efficiency
ratio of the ISR $3\pi$ selection and the $4\pi$ (ISR $4\pi$) selection. 
Events of the ISR 4$\pi$ sample with a generated invariant $4\pi$ mass, $M_{4\pi}^{\rm
truth}$, larger than $3.7 \gev$ are classified as the 
4$\pi$ sample, since they behave quite similar due to the soft ISR photon.

Finally, the efficiency ratio is used to scale the $4\pi$ (ISR $4\pi$) sample selected from data to obtain the background contamination from $4\pi$ (ISR $4\pi$) as a 
function of the 3$\pi$ invariant mass.

\section{Unfolding}\label{s_unf}
The background-subtracted 3$\pi$ mass spectrum needs to be unfolded to account for effects of FSR and detector response before the cross section can be determined. 
The simplified Iterative Dynamic Stable method (IDS)~\cite{c_ids} is used to 
perform the unfolding. It requires the transfer matrix \(A_{ij}\) as input, which correlates the number of actually
produced events in bin \(j\) to the number of reconstructed events in bin \(i\).

Different bin sizes are used in the mass spectra to account for both the mass resolution and the different widths of the resonances: $2.5\mev$ 
in the energy region below $1.05\gev$ (low energy region), and $25\mev$ in the energy region above (high energy region). According to MC studies, the resolution 
varies from $5.2\mev$ to $6.3\mev$ in the low energy region. At high energies it increases to a range from $6.3\mev$ to 
$8.8\mev$ for the tagged and from $5.9\mev$ to $11.0\mev$ for the untagged ISR method. The unfolding is performed 
separately for each energy region. The  $\omega$ and $\phi$ resonances in the low energy region are narrow. To ensure a 
reliable unfolding result, it is important to consider the differences in the detector resolution between MC and data. 
In the high energy region, where both the widths of the resonances ($\omegap$ and $\omegapp$) and the bin sizes are 
much larger than the detector resolution, the differences of the latter between MC and data is negligible to the unfolded spectra. 

To study differences between MC and data, the signals of $\omega$ and
$\phi$ in data are fitted using the shape of 
the 3$\pi$ MC mass distribution to describe the signal. The MC spectrum is convolved with a Gaussian function where the 
parameters reflect possible mass shifts and differences in the
detector resolution between simulation and data. 
Combined results from the fits to $\omega$ and $\phi$ indicate a shift of the masses by $-0.53\pm0.25\mev$, and a 
broadening of the mass resolution by $2.26\pm0.51\mev$ for Data I,
while a slightly larger mass shift of 
$-1.25\pm0.01\mev$ and a negligible modification of the resolution are found for Data II. These numbers are used to 
correct the transfer matrix extracted from simulation, and the errors are considered as systematic uncertainty. 

The stability of the IDS method is provided by using regularization functions~\cite{c_ids}, which avoid unfolding large 
fluctuations in data, \emph{e.g.} due to subtraction of a large background contribution. The parameters of the regularization 
functions need to be optimized for a specific unfolding task.
The optimization is done by first constructing a toy MC distribution from which a `reconstructed toy' is obtained with the transfer matrix from MC.
This toy is made to keep the `reconstructed toy' as close to the real data as possible. The errors for the `reconstructed toy'
are taken directly from  real data and statistically fluctuated, 
including the uncertainty from background subtraction. Lastly, the `reconstructed toy' is unfolded to compare with the toy, and the difference is measured by
a $\chi^2$. A scan for the parameters of the regularization functions is 
performed to find the optimized one with minimum $\chi^2$.

The IDS procedure is applied to the 3$\pi$ mass spectrum obtained after the event selection, the background subtraction, 
and after applying corrections for data/MC efficiency differences (see Sec.~\ref{s_sys} below). It provides the distributions 
unfolded from detector resolutions, distortions, and FSR effects as a function of $\sqrt{s'}$, and a covariance matrix 
of the statistical uncertainties and their bin-to-bin correlations. The $3\pi$ mass spectra after unfolding in 
different mass regions are shown in Sec.~\ref{s_fit_b}.

\section{Systematic uncertainties}\label{s_sys}
The contributions of the tracking efficiency, the photon efficiency, the event selections, and the background 
subtraction to the systematic uncertainties are studied.

\subsection{Tracking efficiency}
The tracking efficiency for pions is studied with tagged $\ee\to\gami2(\pipi)$ events. Each event candidate is required 
to have at least three good charged tracks, where two of them must be oppositely charged. The information on 
$d\mathrm{E}/dx$ and TOF is used to identify the tracks as pions.

The most energetic good photon is required have an energy greater than
$1.2\GeV$ and is assumed to be the ISR photon.  If there is 
more than one good photon, the events with photon pairs yielding invariant masses with $0.12\leq M_{\gami\gamma} \leq 
0.145\gev$ are rejected. Additionally, the energy of the second most energetic photon must be less than $0.07\GeV$.

Events with a second moment of the EMC reconstructed cluster of the
ISR photon is greater than $12\,\text{cm}^2$ and an energy of the selected 
$\gami \pipi \pi^\pm$, $E_{\gam3\pi}$, less than $3.0\GeV$ are
rejected to remove fake ISR photons. According to the MC simulation,
these events come from interactions of $\bar{n}$ particles inside the EMC. The second moment of the shower is defined as 
$\frac{\sum_{i}E_ir_i^2}{\sum_{i}E_i}$, where $r_i$ is the radial distance of the $i^\text{th}$ crystal from the center of 
the cluster, and $E_i$ is the energy deposited in that crystal.

Exploiting the known initial $\ee$ four-momenta and the measured information of the ISR photon, the pair of oppositely 
charged tracks, and one of the other charged tracks,  one can predict the four-momentum of the fourth charged 
track in the event by a $1C$ kinematic fit.

We use the value $\chi^{2}=\chi^{2}_{1C} +\chi^2_{\rm PID}$, where 
$\chi^2_{\rm PID}=\sum\limits_{i=1}\limits^{3}\chi^{2}_{\rm PID}(i)$~\footnote{$\chi^{2}_{{\rm PID}}(i)= 
(\frac{d\mathrm{E}/dx_{\rm measured}-d\mathrm{E}/dx_{\rm expected}}{\sigma_{d\mathrm{E}/dx}})^2+ (\frac{{\rm TOF}_{\rm measured}-{\rm TOF}_{\rm expected}}{\sigma_{\rm TOF}})^2$ for each particle hypothesis $i$ (pion, kaon, and proton).} 
and $\chi^{2}_{1C}$ is the measure of quality from the $1C$ kinematic fit, to identify the charged particles. Only events identified as $\gami2(\pipi)$ 
with $\chi^2_{1C}<5$ and $\chi^2_{\rm PID}<60$ are kept. Any pair of oppositely charged tracks among the three 
selected pion candidates is required to have an invariant mass different from the mass of the $K_S$, \emph{i.e.} not within
\mbox{$0.487 \leq M_{\pi^+\pi^-} \leq 0.507\gev$}.

In the 1C kinematic fit, there is a missing charged track which is used to study the tracking
efficiency. The missing track is taken as being reconstructed in the detector only if
four charged tracks with zero net charge have survived the basic selections in Sec.~\ref{s_sel}. The tracking efficiency is defined as 
\begin{equation}
	  \varepsilon=\frac{2\times N_{ch=4}}{N_{ch=3}+2\times N_{ch=4}},
	\end{equation}
	where $N_{ch=x}$ is the number of events with $x$ charged tracks reconstructed. The factor two in the formula comes from the multiplicity of the track under study.
The contamination to the selected $\ee\to\gami2(\pipi)$ sample 
by residual background contributions is estimated 
using the generic MC samples. The main background contribution comes from $\ee\to(\gami)\piz2(\pipi)$. It is 
determined from the continuum MC sample $\ee\to\qqb$ and compared to the corresponding result of the \babar 
Collaboration~\cite{c_sys_trking_5pi_babar}. A scaling factor of $0.18\pm0.06$ is applied to correct the contribution of 
the process to the generic MC sample. Other background processes contribute to less than 1\%. The MC samples are 
subtracted from data after being scaled according to the integrated luminosity. Effects from background contributions to 
the data-MC efficiency difference are considered by applying a correction $\Delta\varepsilon'$, where $\Delta\varepsilon' = 
\frac{\varepsilon_{\rm Data}^{\rm raw}}{\varepsilon_{\rm MC}}-1$. Here, $\varepsilon^{\rm raw}_{\rm Data}$ is the 
tracking efficiency in data before background (BG) subtraction, and $\varepsilon_{\rm MC}$ is the tracking efficiency in signal MC. 
The difference between MC and data is obtained as:
\begin{linenomath}
 \begin{equation}\label{e_sys_trk_cor}
  \Delta\varepsilon=\frac{\varepsilon}{\varepsilon_\text{MC}}-1=\frac{\Delta\varepsilon'-R_\text{BG}\times\Delta\varepsilon_\text{BG}}{
1-R_\text{BG}},
 \end{equation}
\end{linenomath}
where $\varepsilon$ is the tracking efficiency of data after background subtraction, $R_\text{BG}$ is the background ratio, 
$\varepsilon_\text{BG}$ is the tracking efficiency obtained with the generic MC samples, and $\Delta\varepsilon_\text{BG}=\frac{\varepsilon_\text{BG}}{\varepsilon_\text{MC}}-1$ is the efficiency difference between background and signal. The systematic uncertainties of the 
tracking efficiency due to the event selection criteria are considered by varying the selection criteria within a small range. 
The differences in the results are calculated as a function of both transverse momentum $P_T$ and polar angle 
$\cos\theta$. The largest deviations are taken as the corresponding systematic uncertainties, which, in total, are less 
than 0.2\%. 

Excellent agreement in the tracking efficiency is observed for $\pip$ and $\pim$ independently of the data sets. Thus, 
the results of both data sets and for $\pip$ and $\pim$ are combined to minimize the statistical uncertainties. A 
systematic uncertainty of 0.2\% is assigned to each charged track with transverse momentum larger than $0.25\gevc$. For 
tracks with $P_T<0.25\gevc$, corrections are applied based on the above results and the uncertainties of the corrections 
are taken as systematic uncertainties. For $P_T\in(0.05,0.15)\gevc$, a correction of $(-2.5\pm0.5)$\% is applied for 
$\pip$ and $(-2.3\pm1.5)\%$ for $\pim$. For $P_T\in(0.15,0.25)\gevc$, a correction of $(-0.9\pm0.7)$\% is used for $\pip$ 
and $(-0.2\pm0.6)\%$ for $\pim$.

\subsection{Photon detection efficiency}
The photon efficiency is studied using the process $\ee\to\pipi\piz\piz$. The strategy is to tag $\pip$, $\pim$, 
$\piz_{\rm tag}$, and one photon of the untagged $\piz_{\rm untag}$. The other photon of the $\piz_{\rm untag}$ decay 
can be predicted with the initial four-momenta of $\ee$. In each event, there are two $\piz_{\rm tag}$ candidates in the final state and
each $\piz_{\rm tag}$ decays into two photons. As a result, there will be at most two entries if only one $\piz_{\rm tag}$ is found, and at 
most four entries if two $\piz_{\rm tag}$ candidates are found. Exactly two good charged tracks with zero net charge are required, each 
with an energy deposited in the EMC of less than $1.6\GeV$. The $\chi^2_{\rm PID}$ of the $\pipi$ hypothesis should be 
less than 10 and, additionally, less than the $\chi^2_{\rm PID}$ value for other hypotheses (\emph{e.g.} $\kk$ or $\ppb$). There 
must be at least three good photons, from which the best one or two $\piz_{\rm tag}$ candidates are reconstructed with a  $\chi^2_{1C}<10$
from the $1C$ ($\piz$ mass) kinematic fit. In addition, a unique assignment of the 
photons to all the $\piz_{\rm tag}$ candidates is required. The mass of $\pipi\piz_{\rm tag}$ is required to be less 
than $3.2\gev$ to reduce the contribution of very soft $\piz_{\rm untag}$. The two charged tracks and one $\piz_{\rm 
tag}$ are subjected to a kinematic fit, which is performed to select one good photon and to predict the second photon 
of the $\piz_{\rm untag}$ decay. In this $2C$ fit, the four-momenta of all tracks, including the missing photon, are 
constrained to the initial four-momenta of the $\ee$ system and a mass constraint applied to the $\piz_{\rm tag}$ 
candidate. If more than three photons are found in an event, all possible combinations of $\piz_{\rm tag}$ and of the predicted 
photons are considered.  A signal region ($0.115<M_{\gam\gam}<0.155 \gev$) and side band 
regions ($0.05<M_{\gam\gam}<0.09 \gev$ and $0.18<M_{\gam\gam}<0.22 \gev$) in the invariant mass of $\piz_{\rm untag}$ are studied
to further remove background contributions to the process $\ee\to\pipi\piz\piz$. The predicted 
photons are matched with the remaining good photons by comparing the returned values 
($\delta$, ranging from 0 to 1 to describe the similarity) of the \mbox{\sc howNear} function in the HepLorentzVector class~\cite{c_clhep}, where $\delta=\sqrt{\frac{|\vec{p}_1-\vec{p}_2|^2+(E_1-E_2)^2}{\vec{p}_1\cdot\vec{p}_2+1/4\times(E_1+E_2)^2}}$ defines the difference between two vectors $p_1(E_1,\vec{p}_1)$ and $p_2(E_2,\vec{p}_2)$. Due to the 
resolution difference, requirements are applied as: $\delta_{\rm b}<0.3$ for photons in the barrel region, and 
$\delta_{\rm e}<0.6$ for photons in the end cap of the EMC.

\begin{sloppypar}
The background contamination to the $\ee\to\pipi\piz\piz$ sample is estimated using the
 $\ee\to\qqb$ MC sample, which is modified by substituting the $\ee\to\gami\pipi\piz$ events with $\ee\to\gami\pipi\piz$ in {\sc phokhara}~\cite{c_phokhara}. The background contributions estimated with the generic MC samples is about 1\% 
in the data sample. The photon efficiency is defined as the ratio of the number of entries satisfying the requirement on 
$\delta$ to all the entries. The raw efficiency of the data is corrected in the same way as done for the tracking 
efficiency. The efficiency difference between MC and data is assumed to be constant with respect to the photon energy 
($E$) and the polar angle ($\cos\theta$). The average values are obtained through fitting, and the largest one is taken 
as the systematic uncertainty of the photon efficiency.\end{sloppypar}

In order to estimate the corresponding systematic uncertainties, all criteria used to select the control sample are 
varied by 3\% for $M_{3\pi}$, by 10\% for $E^{\rm EMC}$, and by at least 20\% for others. The resulting uncertainty is found to be $(-0.17\pm0.05\pm0.06$)\%. To be conservative, to each photon is assigned 
a systematic uncertainty of 0.3\%.

\subsection{Uncertainties  from the event selection criteria}
The event selection criteria determine its efficiency. Understanding the systematic uncertainties of the 
criteria can be achieved through studies of control samples. These samples can be constructed from data 
using criteria similar to the ISR 3$\pi$ event selection discussed above.
Since the selection criterion under study is not used, the purity of these control 
samples is guaranteed by choosing the $\omega$ and $\jpsi$ signals in the 3$\pi$ mass
spectra and by tightening other selection criteria.

For the tagged ISR method, the control samples are selected from the tagged ISR 3$\pi$ signal with mass windows 
($0.75<M_{\pipi\piz}<0.82 \gev$) for $\omega$ and ($3.08<M_{\pipi\piz}<3.12 \gev$) for $\jpsi$. The bias is assumed to be linear in the 
3$\pi$ mass and can be obtained for a given 3$\pi$ mass by fitting. The main background in these two samples is $4\pi$ (ISR 
$4\pi$), which has been measured in Sec.~\ref{s_bg}. Thus, the $4\pi$ (ISR $4\pi$) MC is scaled to data according to the $3\pi$ 
mass spectrum in the $\omega$ and the $\jpsi$ mass windows, respectively. The contribution of other background processes
is small and can be estimated with the generic MC sample. The signal efficiency is compared to the value 
obtained from the control samples after subtracting background contributions.

For the untagged ISR method, the control sample is selected from the untagged ISR 3$\pi$ signal with 3$\pi$ mass window 
($3.07<M_{\pipi\piz}<3.13 \gev$) for $\jpsi$. The background contamination is much smaller compared to the tagged control samples. 
Remaining background contributions are removed analogously to the tagged control samples. Since only one single control 
sample is used, the bias is assumed to be constant from 1.05 to $3\gev$.

To reduce the background contributions in the control 
samples, some requirements are further tightened: The charged tracks must be identified as pions by means of $\chi^2_{\rm PID}$; the penetration depth of tracks in the MUC is required to be less than 40\,cm to 
reject muons in the $\jpsi$ control samples.

For the evaluation of the condition on $E_{\rm EMC}/p$, the tagged control samples are constructed by requiring 
$\chi^2_{5C}<25$ and $\chi^2_{1C} (\piz)<5$. The biases are determined to be $(-0.57\pm0.39)\%$, 
$(-0.21\pm0.40)\%$, and $(-0.07\pm0.24)\%$ by comparing the efficiencies from signal MC and control samples in data in 
the mass regions of $\omega$ and $\jpsi$ for the tagged analysis, and the $\jpsi$ region for the untagged analysis. 
The 
quoted errors include both the statistical and the systematic error due to the uncertainty of scaling the $4\pi$ (ISR $4\pi$) 
background.

The two photons ($\gam^1_{\piz}$, $\gam^2_{\piz}$) in the final state are constrained to the nominal $\piz$ mass, and 
$\chi^2_{1C}<10$ is required. This is applied on top of the $\fCtC$ kinematic fit and the requirement 
$\chi^2_{\fCtC}<60\,(25)$. Therefore, the bias of the initial requirement is negligible. However, side bands to 
the $\piz$ peak are used to remove background contributions and this must be considered. Three control samples are 
selected analogously to the $E_{\rm EMC}/p$ study. By comparing the $\chi^2_{1C}$ distributions after the subtraction of 
the $\piz$ side bands, biases from $\chi^2_{1C}<10$ and the $\piz$ side bands subtraction are determined to 
be $(0.8\pm0.5)\%$, $(1.2\pm0.5)\%$, and $(-0.1\pm0.3)\%$ from the three control samples.

A kinematic fit is used in both ISR methods. Its bias is evaluated for the three data samples. To reduce 
background contamination, on top of the standard selection criteria, the energy of the tagged ISR photon is required to 
be $E_{\gami}^{\omega}>1.72\GeV$ and $0.55<E_{\gami}^{\jpsi}<0.65\GeV$, while the condition on $\chi^2_{\fCtC}$ is 
dropped. For the untagged sample, only events with two good photons are considered. Additionally, the $\jpsi$ side 
bands ($2.92<M_{\pipi\piz}<2.98 \gev$ and $3.20<M_{\pipi\piz}<3.26 \gev$) are used to subtract background contributions. By comparing the 
$\chi^2_{\fCtC}$ distributions, biases from $\chi^2_{\fCtC}<60\,(25)$ are determined as $(-0.36\pm0.58)\%$,
$(0.67\pm0.58)\%$, and $(0.75\pm0.23)\%$ from the tagged $\omega$, tagged $\jpsi$, and untagged $\jpsi$ control samples.

The additional $\piz$ veto in the tagged ISR method is studied by applying more stringent requirements on the 
$\chi^2$ values: $\chi^2_{5C}<25$ and $\chi^2_{1C} (\piz)<5$.  A value of $(-0.10\pm0.35)\%$ is assigned to the tagged $\omega$ 
sample, and a correction of $(0.83\pm0.74)\%$ for the tagged $\jpsi$ sample.

The bias due to the condition on $|\cos\theta|$ of the untagged ISR photon is estimated from a comparison of 
the angular distributions in MC and data using the untagged $\jpsi$ sample. The relative efficiency
difference between the MC and the data is taken as the systematic uncertainty for the selection $|\cos\theta|>0.9984$. 
The  results of the two data sets are 
combined yielding $(-0.64\pm1.49)\%$ as the systematic uncertainty. The main source comes from the uncertainty of the
background subtraction.

The vertex position given by the vertex fit is used to reject background contributions. 
Using a $\ee\to\gami\jpsi$  sample, the efficiency difference between MC and data due to the vertex fit is estimated to be less than 0.2\%.

\subsection{Systematic uncertainty from background subtraction}

\begin{figure*}
	\begin{center}
		\includegraphics[width=0.33\textwidth]{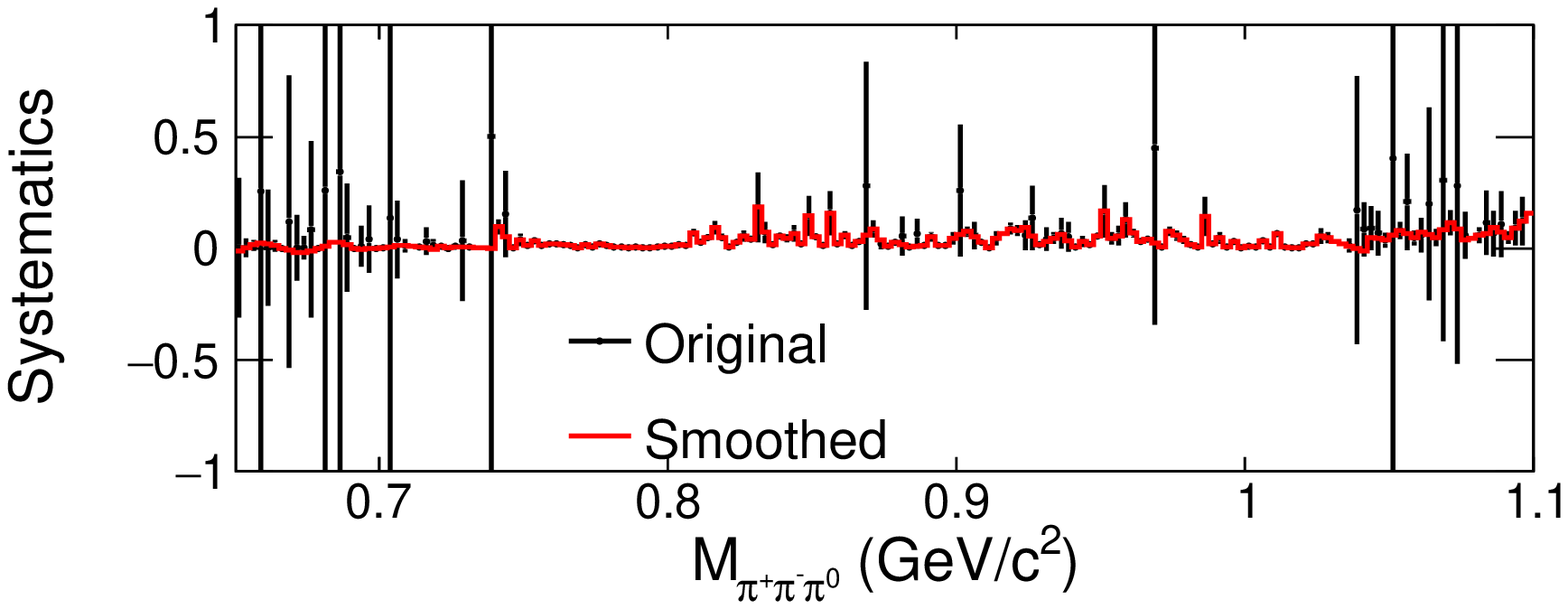}%
		\includegraphics[width=0.33\textwidth]{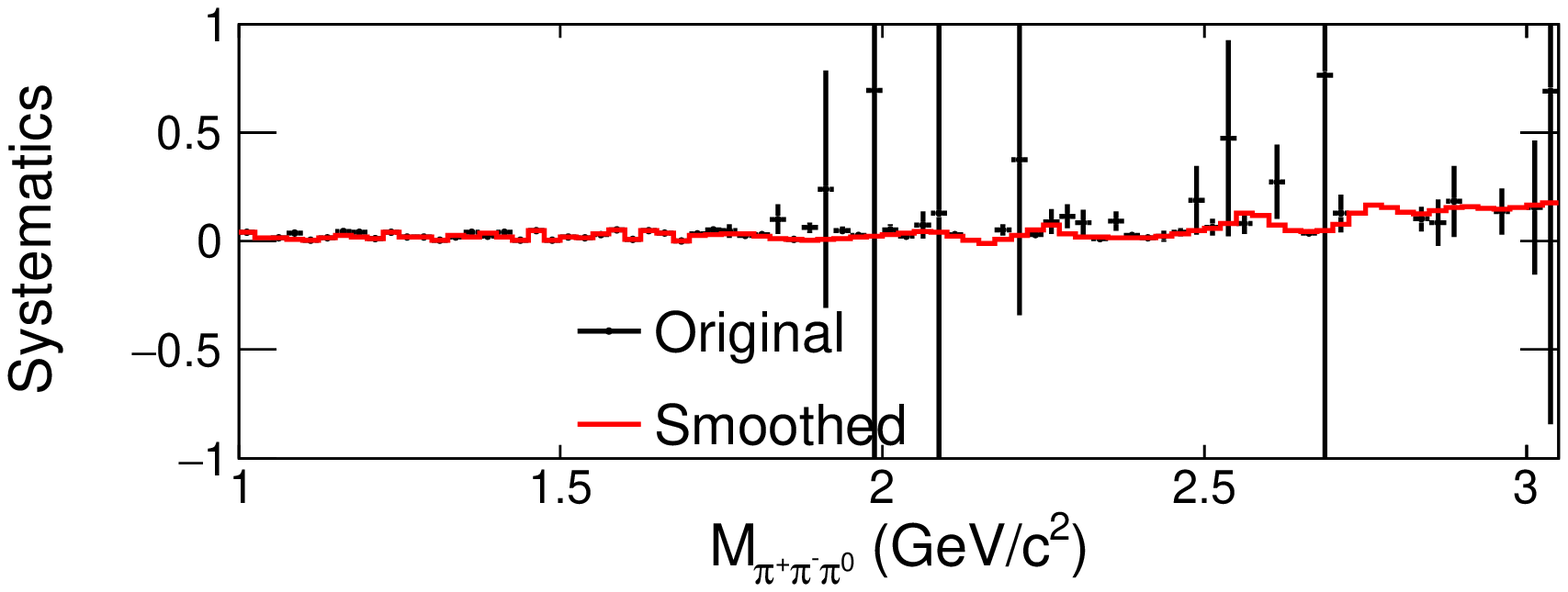}%
		\includegraphics[width=0.33\textwidth]{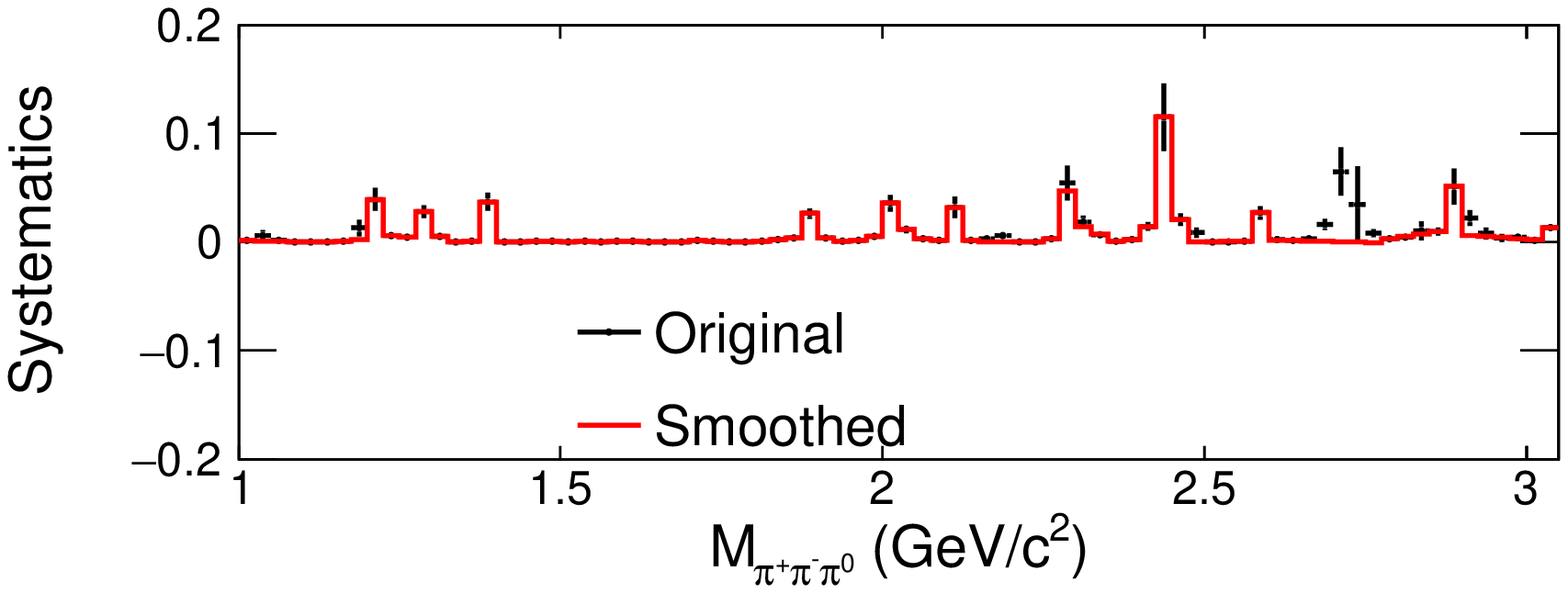}\\%
		\includegraphics[width=0.33\textwidth]{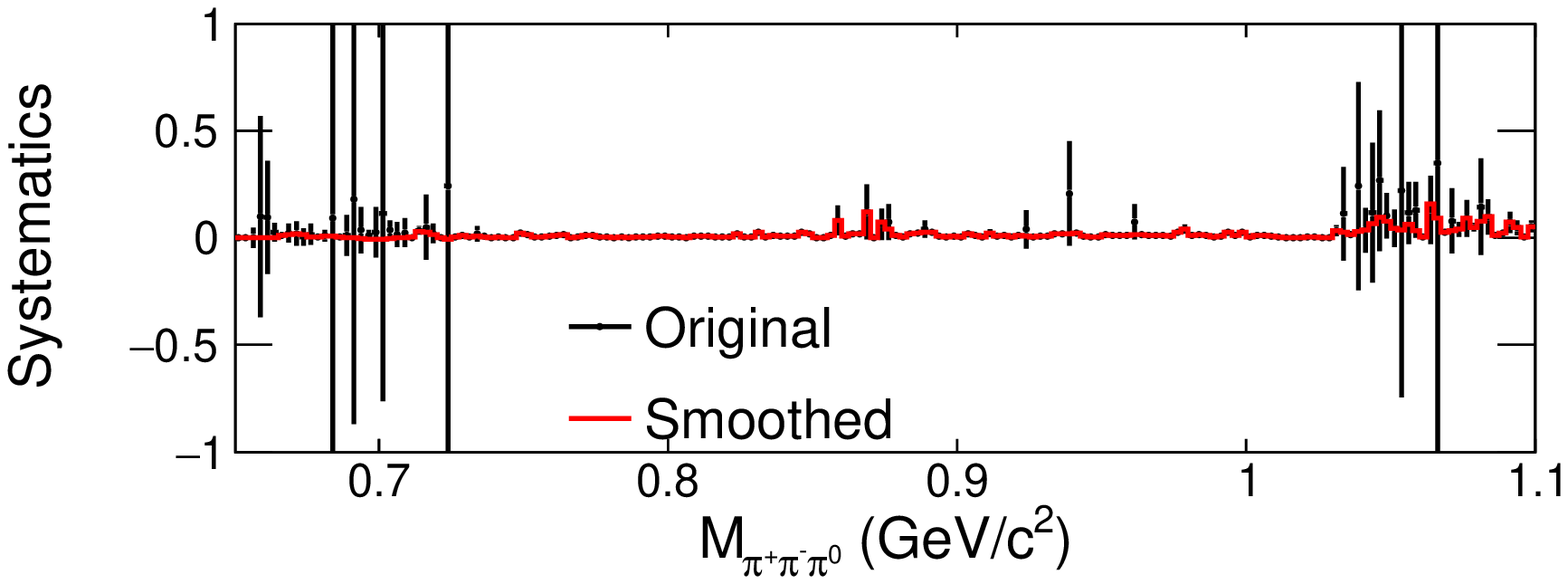}%
		\includegraphics[width=0.33\textwidth]{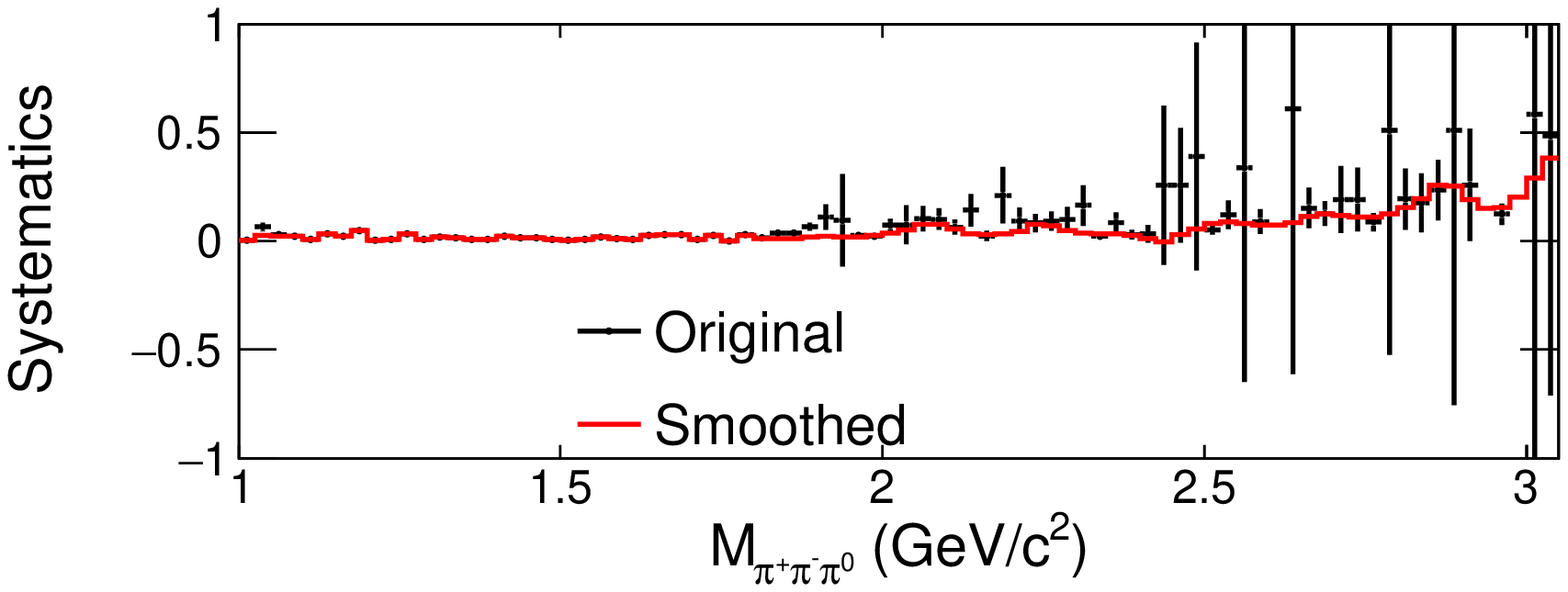}%
		\includegraphics[width=0.33\textwidth]{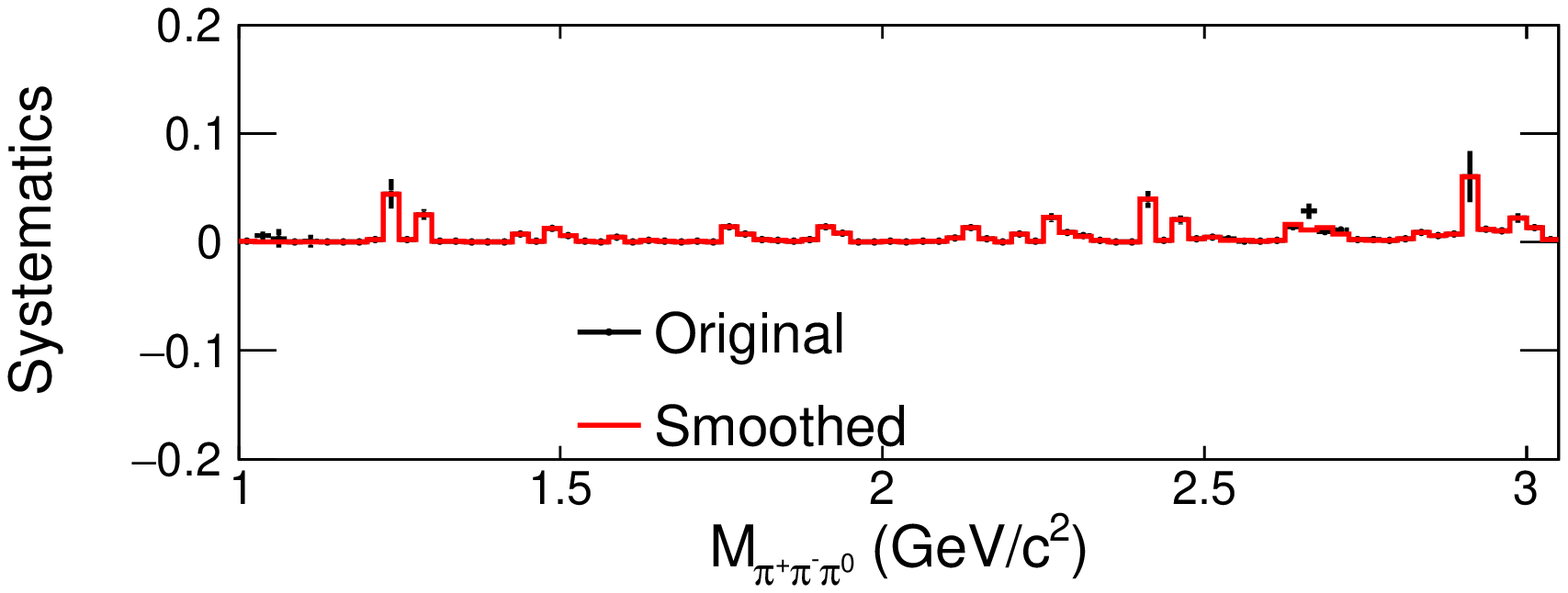}%
		\caption{\label{f_sys_subBG}Systematic uncertainties for background subtraction as a function of $M_{\pipi\piz}$. 
		The black points are the original numbers, the red curves are the smoothed results; the left and the middle plots are for tagged ISR selection, the right ones are for untagged ISR selection; the top row is for Data I, and the bottom row is for Data II.}
	\end{center}
\end{figure*}

The main background contributions are $\ee\to(\gami)\pipi\piz\piz$. Its uncertainty is dominated by the statistical 
uncertainty of the selected data samples, which are scaled by the efficiency ratio obtained from MC. Remaining 
systematic uncertainties are estimated by varying the event selection criteria, mainly the $\chi^2$ from kinematic fit, 
by at least 40\%. The maximum difference of the unfolded mass spectra between the standard selection and the variation is taken as a measure for the 
systematic uncertainty. Other background contributions are subtracted with the generic MC samples. Further sources of 
systematic uncertainties stem from branching fractions, cross sections, and dileptonic widths 
$\Gamma_{\ee}^{\psip/\jpsi}$~\cite{c_pdg2014}. The systematic uncertainty of the backgound subtraction is estimated by 
varying all estimated background contributions by 1$\sigma$. The unfolded mass spectra are compared bin-by-bin after 
subtracting the background, and the maximum relative difference is taken as the systematic uncertainty in the corresponding bin. 
A smoothing procedure is applied to reduce the large statistical errors in a few bins with poor
statistics. The resulting systematic uncertainty 
 of the background subtraction is shown in Fig.~\ref{f_sys_subBG} and varies from 0.3\% to 26\% for the tagged ISR method and from 0\% to 12\% for the untagged ISR method. 

\subsection{Systematic uncertainty from unfolding}

The transfer matrix is corrected due to a mass shift and a resolution difference between MC and data. The corrections 
to the transfer matrix are varied by 1$\sigma$ to study the systematics. Additionally, tests of the unfolding procedure 
are performed to investigate potential biases introduced by the method. Toy experiments are carried out similarly to the ones described in Sec.~\ref{s_unf}, except for the statistical fluctuation of the `reconstructed toy'. The approach 
allows studying the actual systematics of the unfolding method, avoiding bias from the statistical uncertainty of the 
data sets. Comparisons are made between the `unfolded toy' and the true toy. The relative difference is taken as 
the systematic uncertainty. To reduce large statistical effects of individual bins with only a few events, an averaged 
value is assigned to each bin by fitting the difference with a constant. The systematic uncertainties are estimated to 
be 0.7\% below $1.05\gev$ and  1.2\% above $1.05\gev$ for tagged, and 1\% for the untagged mass spectrum.

\subsection{Summary of the systematic uncertainties}
Biases are corrected for and the uncertainties of the corrections are taken as the remaining systematic 
uncertainties for the cross section. The individual contributions to the systematic uncertainties are summarized in Table~\ref{t_sum_sys}. 
Details about the vacuum polarization can be found in
Sec.~\ref{s_res}. The total numbers for the different mass regions are 
also listed. The average uncertainty in a certain mass range is calculated by weighting the ratio of events in each 
bin from signal MC.
\begin{table*}[!hbt]
 \caption{\label{t_sum_sys}Summary of the relative systematic uncertainties for the $\ee\to\pipi\piz$ cross section (in \%). The total systematics are summarized 
below for the different mass regions.}
 \begin{center}
  \begin{tabular*}{\textwidth}{@{\extracolsep{\fill}}l|c|c|c|c}
  \hline\hline
  Data samples &\multicolumn{2}{c|}{Data I}&\multicolumn{2}{c}{Data II}\\
  \hline
  Source  & Tagged & Untagged &Tagged & Untagged\\
  \hline
  Tracking                               & 0.4-1.0 & 0.4-0.9 & 0.4-0.7 & 0.4-0.7 \\ 
  Photon reconstruction                  &  0.9   &  0.6   &  0.9   &  0.6   \\ 
  E/P                                    & 0.7-0.9 &  0.4   &  0.4   &  0.4   \\ 
  $\piz$ side band                       & 0.6-0.9 &  0.4   &  0.4   &  0.4   \\ 
  Kinematic fit ($\chi^2$ cut)           & 1.0-1.4 &  0.4   &  0.6   &  0.4   \\ 
  Veto $\piz$ for $\gam_{\rm ISR}$       &  0.6   &   -    & 0.5-1.0 &   -    \\
  $\cos{\theta_{\gam_{\rm ISR}}^{2C}}$   &   -    &  1.5   &   -    &  1.5   \\
  Vertex                                 &   -    &  0.2   &   -    &  0.2   \\
  BG subtraction                         & 0.0-19.0& 0-12.0  &0.04-26.0& 0.0-6.1 \\
  Vacuum polarization                    & 0.1-0.3 &  0.1   & 0.1-0.3 &  0.1   \\
  Unfolding                              & 1.0-1.7 &  1.3   & 0.7-1.0 &  0.8   \\
  Radiative function                     &  0.5   &  0.5   &  0.5   &  0.5   \\
  Luminosity                             &  1.1   &  1.1   &  1.1   &  1.1   \\ 
  \hline
  Total                                  & $>2.4$  & $>2.6$  & $>2.0$  & $>2.4$  \\
  \hline
  ${\rm Total}_{\rm low}$ (0.7-1.05)[average]   &2.1-16.0[2.6]&  -        &1.8-11.0[2.3]&  -        \\
  ${\rm Total}_{\rm medium}$ (1.05-2.0)[average]&2.4-5.9[3.5] &2.0-4.4[2.1]&1.9-7.4[2.7] &1.6-4.8[1.7]\\
  ${\rm Total}_{\rm high}$ (2.0-3.0)[average]   &2.8-19.0[8.3]&1.9-12[2.5] &3.1-25.0[11] &1.6-6.3[2.0]\\
  \hline
  ${\rm Total}_{\omega}$ (0.76-0.8)[average]    &2.2-3.0[2.3]&  -        &1.8-6.5[2.1]&  -        \\
  ${\rm Total}_{\phi}$ (1.01-1.03)[average]     &2.1-7.0[2.7]&  -        &1.8-5.1[2.0]&  -        \\
  ${\rm Total}_{\omegap}$ (1.1-1.4)[average]    &2.5-4.7[3.3]&2.0-4.4[2.6]&1.9-5.3[2.7]&1.6-4.8[2.0]\\
  ${\rm Total}_{\omegapp}$ (1.40-1.80)[average] &2.4-5.9[3.8]&2.0-2.0[2.0]&1.9-3.6[2.6]&1.6-2.1[1.7]\\
  ${\rm Total}_{\jpsi}$ (3.05-3.15)[average]    &  -        &1.4-2.2[1.4]&  -        &1.4-7.4[1.4]\\
  \hline\hline
  \end{tabular*}
 \end{center}
\end{table*}

\section{Combined fit of the 3$\pi$ mass spectrum}\label{s_fit_b}
\begin{figure*}
 \begin{center}
  \subfigure[Tagged low mass region (data I).]{\label{f_fit_m3pi:a}
  \includegraphics[width=0.33\textwidth]{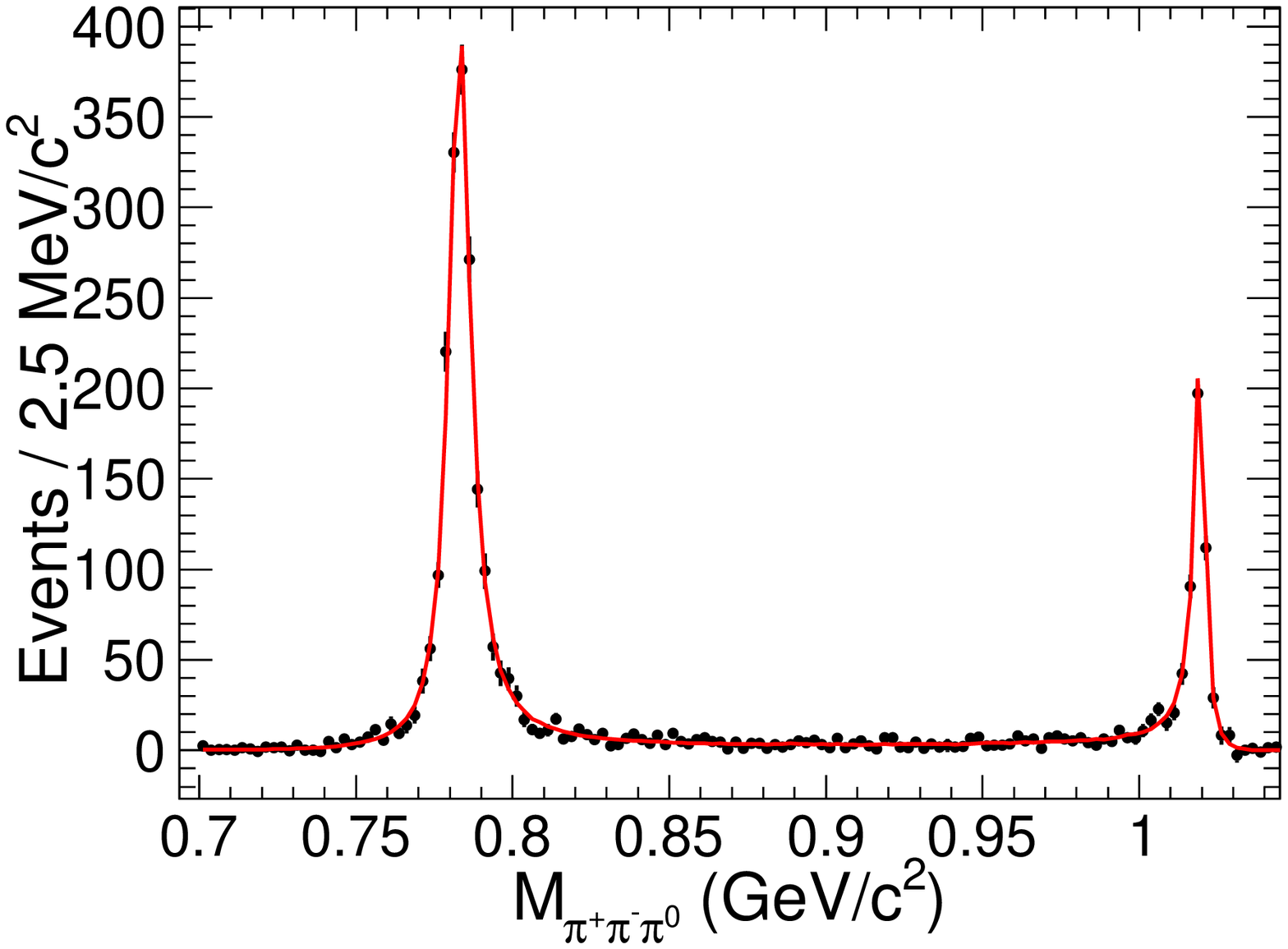}}%
  \subfigure[Tagged medium mass region (data I).]{\label{f_fit_m3pi:b}
  \includegraphics[width=0.33\textwidth]{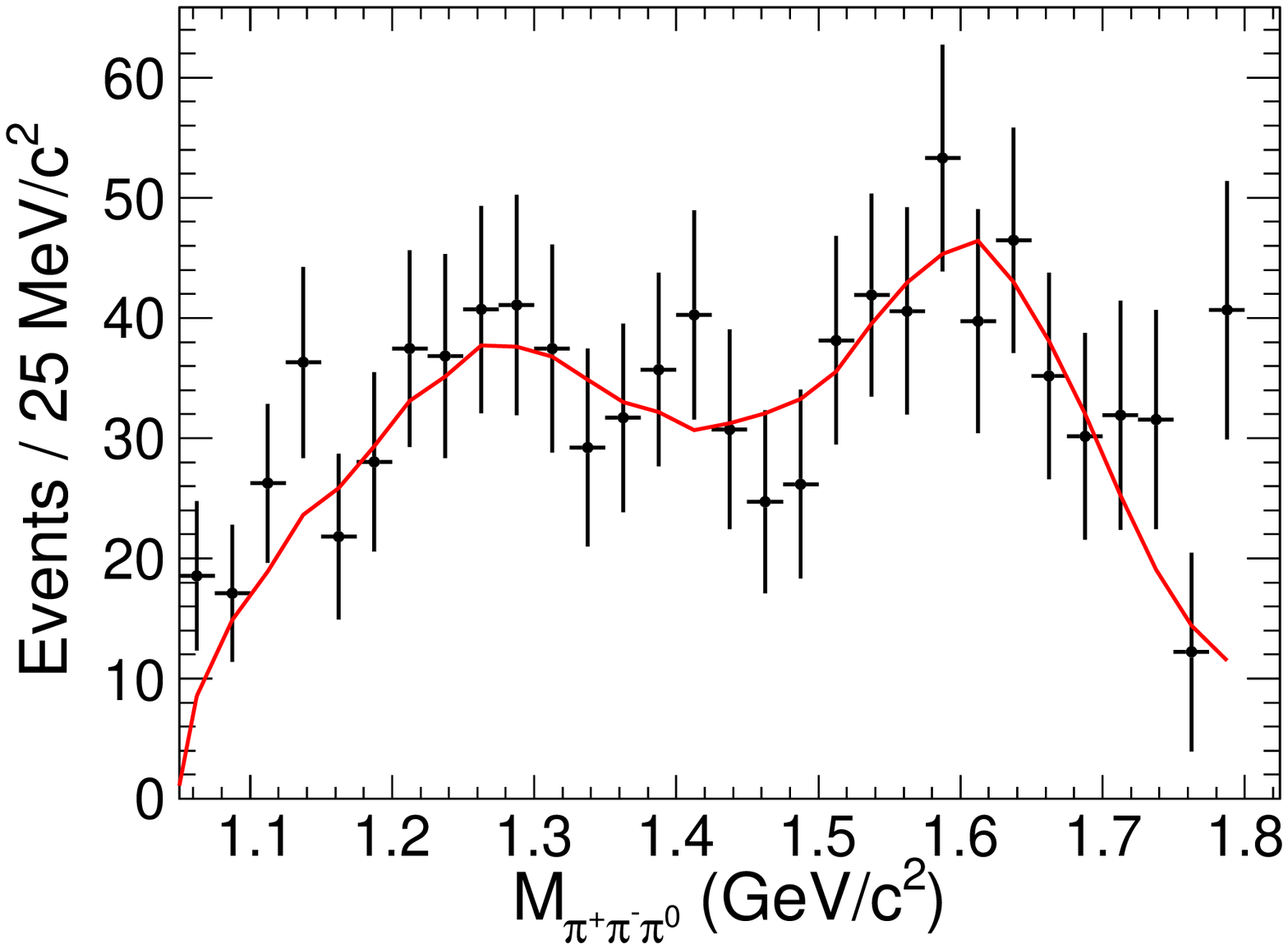}}%
  \subfigure[Untagged  (data I).]{\label{f_fit_m3pi:c}
  \includegraphics[width=0.33\textwidth]{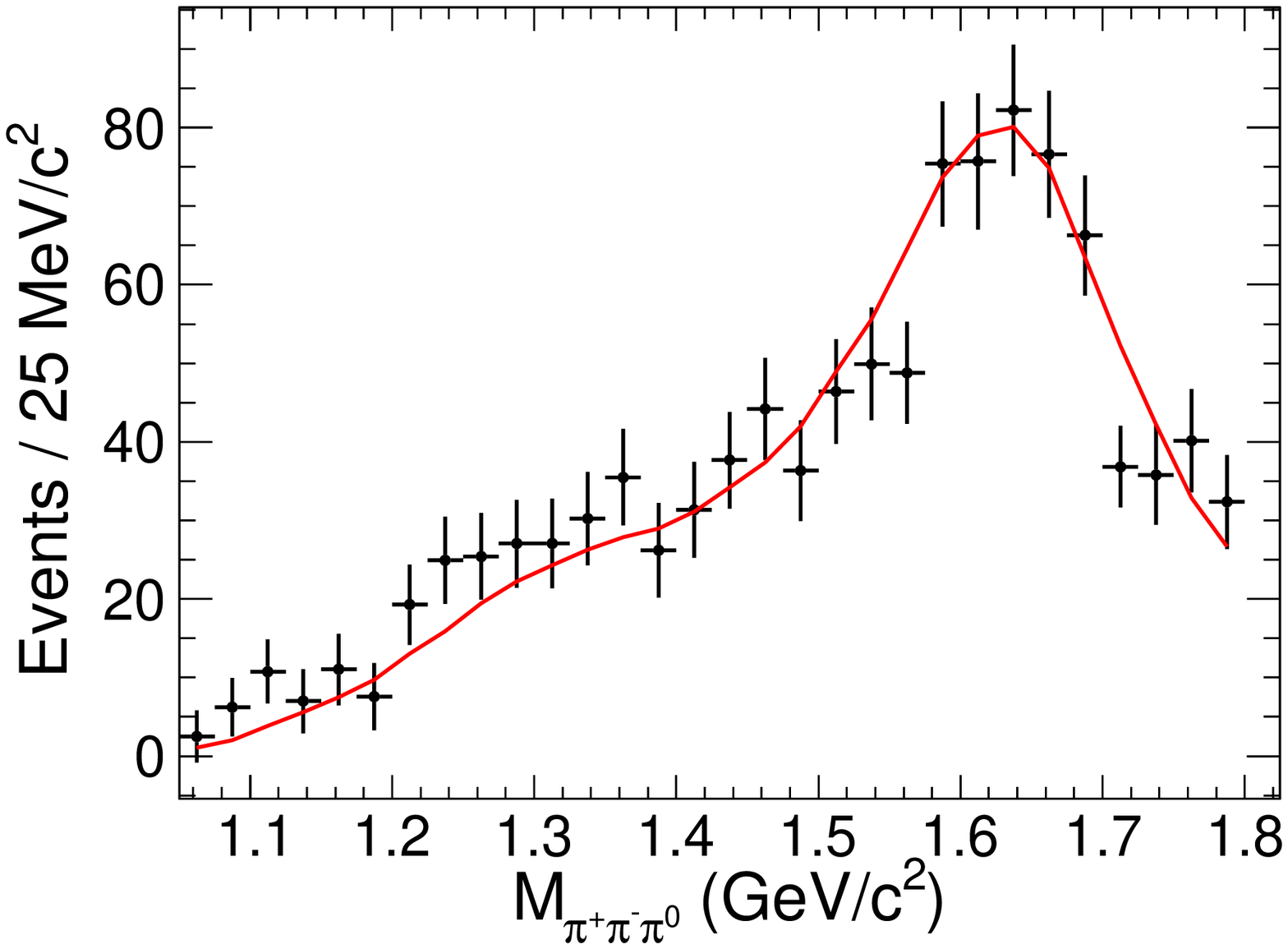}}\\%
  \subfigure[Tagged low mass region (data II).]{\label{f_fit_m3pi:d}
  \includegraphics[width=0.33\textwidth]{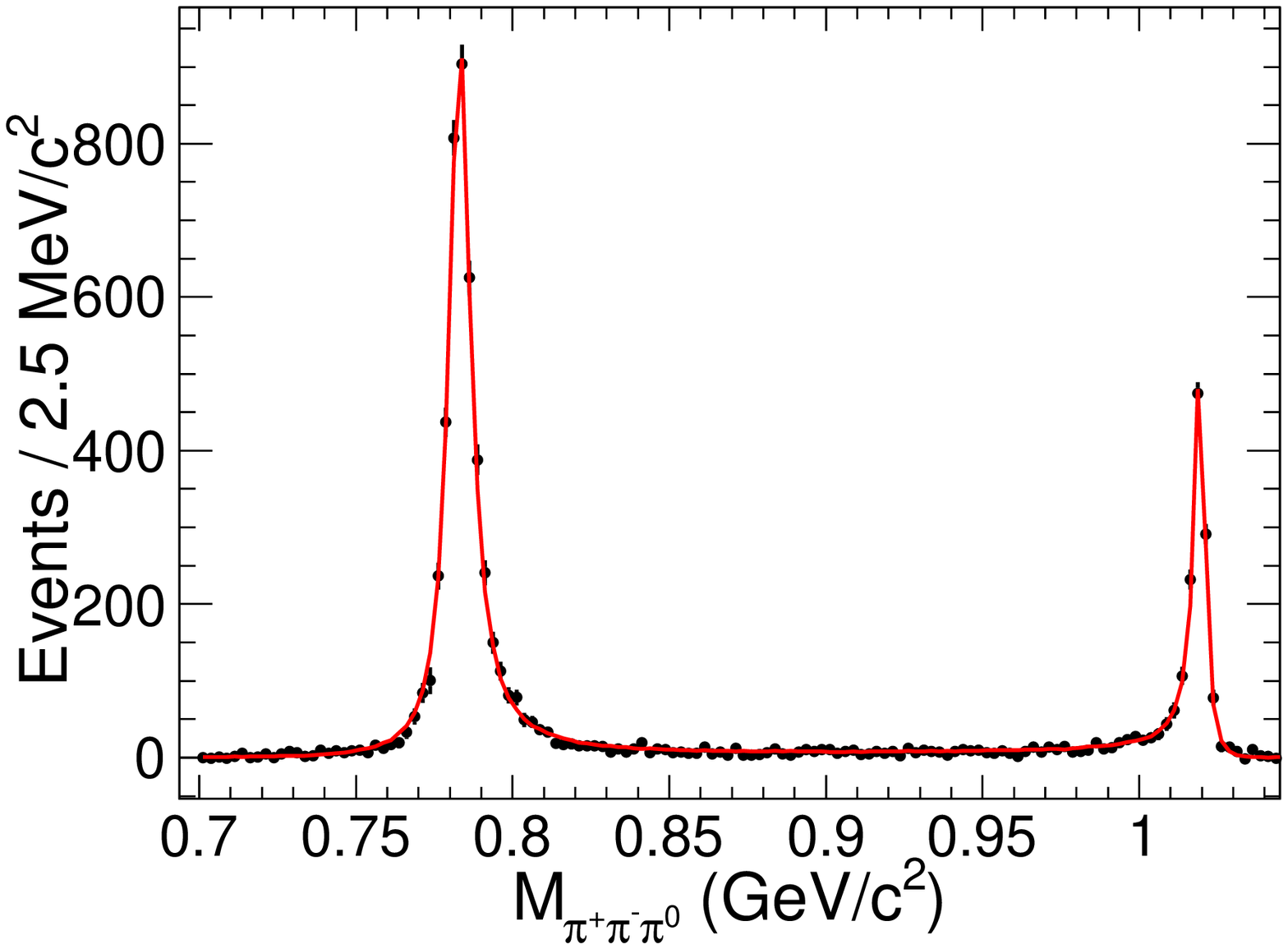}}%
  \subfigure[Tagged medium mass region (data II).]{\label{f_fit_m3pi:e}
  \includegraphics[width=0.33\textwidth]{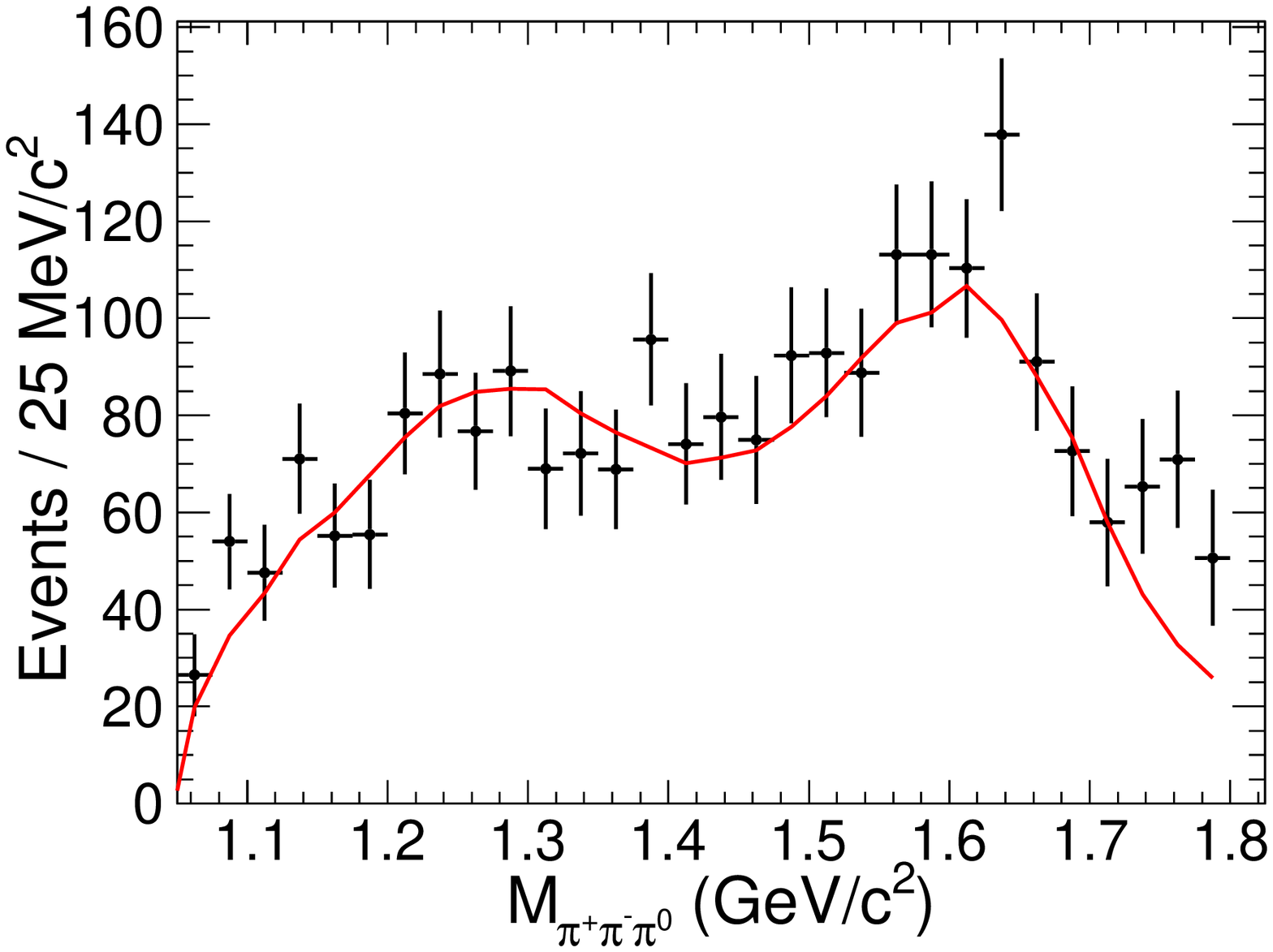}}%
  \subfigure[Untagged  (data II).]{\label{f_fit_m3pi:f}
  \includegraphics[width=0.33\textwidth]{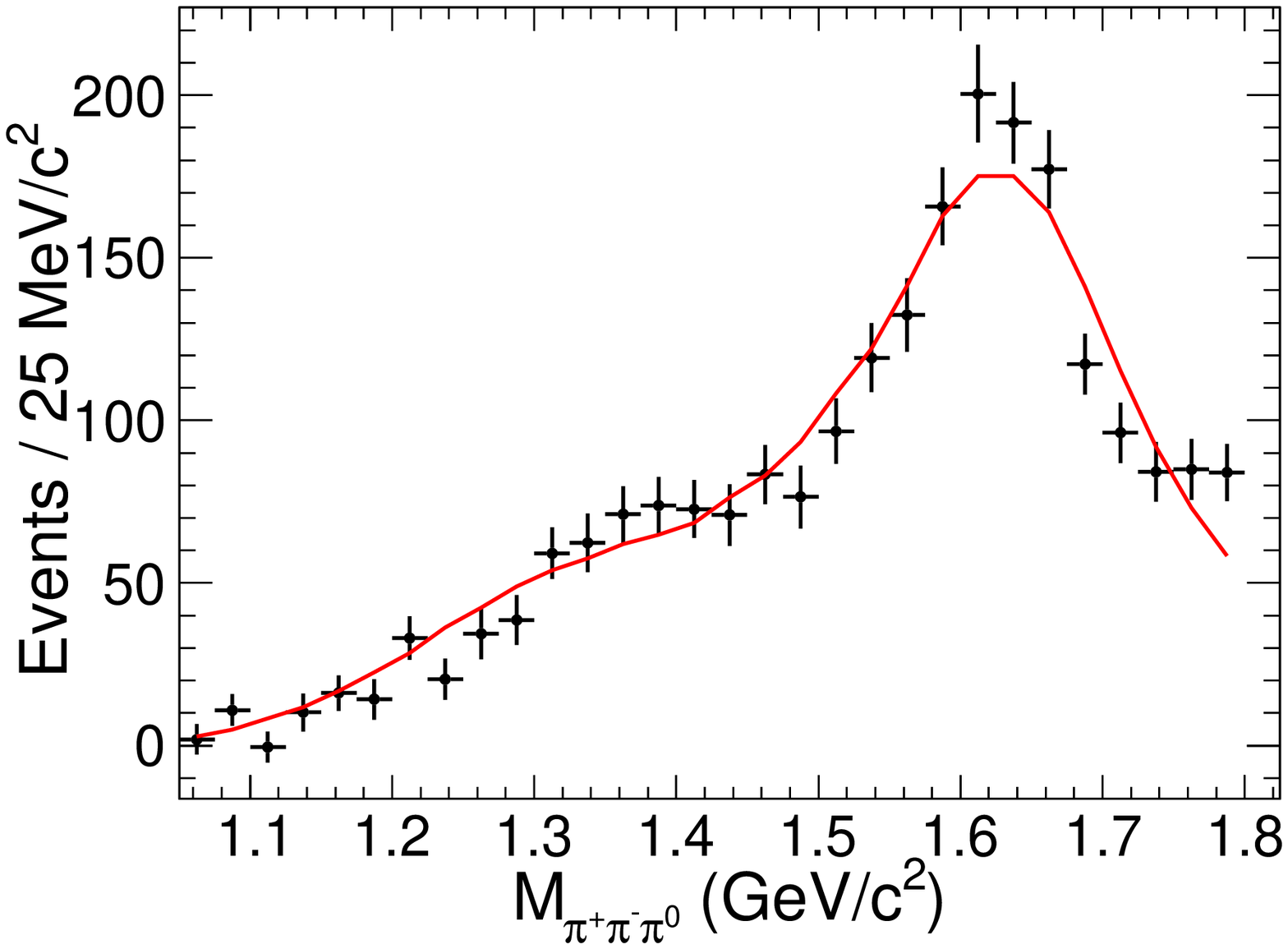}}\\%
 \end{center}
 \caption{\label{f_fit_m3pi}Results of the fit to different mass regions and different data sets.}
\end{figure*}

The parameters of the vector resonances $\omega, \phi, \omegap$, and $\omegapp$ are determined with a combined fit to 
the $3\pi$ mass spectrum from both the tagged and untagged ISR method in both data sets. The relation
\begin{linenomath}
 \begin{equation}
  \label{e_fit_tag}
  \frac{dN}{d\sqrt{s'}}=\varepsilon(\sqrt{s'})\cdot\frac{dL}{d\sqrt{s'}}\cdot\sigma(\sqrt{s'})
 \end{equation}
\end{linenomath}
is fitted to the unfolded mass spectrum between 0.7 to $1.8\gev$, where $\varepsilon$ is the detection efficiency, 
$dL/d\sqrt{s'}$ is the effective luminosity, and $\sigma$ is the Born 
cross section for $\ee\to3\pi$. The effective luminosity, defined as $2\cdot \sqrt{s'}/s\cdot F(s,\sqrt{s'})\cdot L_{ee}$, relies on the total integrated luminosity $L_{ee}$ and the 
calculated radiator function $F(s,\sqrt{s'})$~\cite{c_kuraev}. The latter describes the probability to radiate an ISR 
photon so that the production of hadronic final states with a mass of $\sqrt{s'}$ is possible. According to the VMD 
model, $\sigma({\sqrt{s'}})$ can be written as the sum of four resonances:
\begin{widetext}
 \begin{equation}
  \label{e_fit}
	{\large \sigma(\sqrt{s'})=\frac{12\pi}{s'^{3/2}}F_{\rho\pi}(\sqrt{s'})\left|\sum_{V=\omega,\phi,\omegap,\omegapp}
  \frac{\Gamma_V 
m_V^{3/2}\sqrt{\BR(V\to\ee)\BR(V\to3\pi)}}{m^2_V-s'-i\sqrt{s'}\Gamma_V(\sqrt{s'})}\frac{e^{i\phi_V}}{\sqrt{F_{\rho\pi}(m_V)}}\right|^2 },
 \end{equation}
\end{widetext}
where $m_V$ and $\Gamma_V$ are the mass and width of the vector meson $V, \phi_V$ is the corresponding phase, 
$\BR(V\to\ee)$ and $\BR(V\to3\pi)$ are the branching fractions of $V$ decaying into $\ee$ and $\pipi\piz$, 
respectively, and  $\Gamma_V(m)=\sum_{i}\Gamma_i(m)$ is the total width. Here, $\Gamma_i(m)$ is the 
partial width of the resonance decay into the final state $i$.
The 
mass-dependent widths of $\omega$ and $\phi$ are calculated taking into account all significant decay modes. 
The decay $V\to3\pi$ is assumed to proceed via the $\rho\pi$ intermediate state, and $F_{\rho\pi}(m)$ is the 3$\pi$ phase space 
volume calculated under this hypothesis. Details about the formulas can be found in Ref.~\cite{c_snd25}. A combined 
binned $\chi^2$ fit is performed. The results are illustrated in Fig.~\ref{f_fit_m3pi}. 
The mass of $\omega$ and $\phi$, the mass and width of $\omegap$ and 
$\omegapp$, and the branching fractions are free parameters. The relative phase between $\omega$ and $\phi$ is taken from 
Ref.~\cite{c_snd25}, $\phi_{\phi}=(163\pm7)^{\circ}$, while the phases of $\omega$, $\omegap$, and $\omegapp$ are fixed 
at $0^{\circ}$, $180^{\circ}$, and $0^{\circ}$, respectively~\cite{c_phase30}. The widths of $\omega$ and $\phi$ are fixed to the 
PDG~\cite{c_pdg2014} values.
Table~\ref{t_res_d1} summarizes the results of the fit.

\begin{table*}
 \caption{\label{t_res_d1}Result of the fit to the 3$\pi$ mass spectra, where $\BR_1$ stands for $\BR(V\to\ee)$ and 
$\BR_2$ represents $\BR(V\to3\pi)$. }
 \begin{center}
  \begin{tabular*}{\textwidth}{@{\extracolsep{\fill}}l|c|c|c}
  \hline\hline
	Parameters & PDG~\cite{c_pdg2014} & \babar & This analysis\\
  \hline
  $\chi^2$/NDF&-&146/148&443/390\\
  \hline
  $m_{\omega}~(\zmev)$    & $782.65\pm0.12$  & $782.45\pm0.24$  & $783.20\pm0.07\pm0.24$  \\
  $m_{\phi} ~(\zmev)$     & $1019.46\pm0.02$ & $1018.86\pm0.20$ & $1020.00\pm0.06\pm0.41$ \\
  $m_{\omegap} ~(\zmev)$  & $1400\sim1450$   & $1350\pm20\pm20$ & $1388\pm39\pm55$        \\
  $m_{\omegapp} ~(\zmev)$ & $1670\pm30$      & $1660\pm10\pm2$  & $1699\pm9\pm7$          \\
  \hline
  $\Gamma_{\omega} ~(\mev)$    & $8.49\pm0.08$ & PDG~\cite{c_pdg2014}
  & PDG~\cite{c_pdg2014}\\
  $\Gamma_{\phi} ~(\zmev)$     & $4.25\pm0.02$ & PDG~\cite{c_pdg2014} & PDG~\cite{c_pdg2014}\\
  $\Gamma_{\omegap} ~(\zmev)$  & $180\sim250$  & $450\pm70\pm70$ & $629\pm155\pm221$ \\
  $\Gamma_{\omegapp} ~(\zmev)$ & $315\pm35$    & $230\pm30\pm20$ & $331\pm40\pm29$ \\
  \hline
  $(\BR_1\times\BR_2)(\omega)~(10^{-5})$    & $6.56\pm0.12$ & $6.70\pm0.06\pm0.27$ & $6.94\pm0.08\pm0.16$ \\
  $(\BR_1\times\BR_2)(\phi)~(10^{-5})$      & $4.53\pm0.10$ & $4.30\pm0.08\pm0.21$ & $4.20\pm0.08\pm0.19$ \\
  $(\BR_1\times\BR_2)(\omegap)~(10^{-6})$   & seen & $0.82\pm0.05\pm0.06$ & $0.84\pm0.09\pm0.09$ \\
  $(\BR_1\times\BR_2)(\omegapp) ~(10^{-6})$ & seen & $1.30\pm0.10\pm0.10$ & $1.14\pm0.15\pm0.15$ \\
  \hline\hline
  \end{tabular*}
 \end{center}
\end{table*}
The systematic uncertainties of the fitted masses and widths include contributions from the uncertainties of 
$\Gamma_\omega$, $\Gamma_\phi$, $\phi_{\phi}$, and the corrections of the difference in resolution between MC and data. 
In order to determine these contributions, the fits are performed varying each of the above parameters by $1\sigma$, and the 
differences between the fitted results and the normal fitted result, shown in Table~\ref{t_res_d1}, are taken as the 
corresponding contributions to the systematic uncertainty. The estimated background contributions may also affect the 
fit results of the masses and widths of the resonances. To consider this issue, toy samples are produced for the 
estimated background contributions by randomly fluctuating each bin content according to a Gaussian distribution. Every toy 
sample is subtracted from data and the remaining mass spectrum is fitted. The distribution of 
the obtained parameters is fitted with a Gaussian after 100 fits. Its standard deviation is taken as the uncertainty of the respective 
parameter. Compared to other sources, the uncertainties due to background contamination to the fitted masses and widths 
are negligible.  A similar study is done, except for the background subtraction which has 
been considered in Table~\ref{t_sum_sys}, for the branching fractions. 
To consider the systematics from bin width and fit range, we  fit toy MC samples 
generated with the same model. For each parameter, 100 MC toys are generated, 
and the largest difference from comparing the values of the input and the fitted 
parameter is taken as the systematic uncertainty.
Table~\ref{t_sys_fit} summarizes the total systematic uncertainties from the fit. 
As for the branching fractions, this table reports the additional systematics 
besides Table~\ref{t_sum_sys}.
\begin{table*}
 \begin{center}
   \small{
	 \caption{\label{t_sys_fit}Relative systematic uncertainties
           (Sys., in \%) from the fit. $\BR$ is defined as $\BR(V\to\ee)\times\BR(V\to3\pi)$. }
  \begin{tabular*}{\textwidth}{@{\extracolsep{\fill}}l|c|c|c|c|c|c|c|c|c|c}
   \hline\hline
   Variable &$\BR(\omega)$&$\BR(\phi)$&$\BR(\omegap)$&$\BR(\omegapp)$
   &$M_\omega$&$M_\phi$&$M_{\omegap}$&$M_{\omegapp}$&$\Gamma_{\omegap}$&$\Gamma_{\omegapp}$\\
   \hline
   Sys. (\%)&1.7&4.0&9.7&13&0.03&0.04&3.9&0.4&35&8.5\\

   \hline\hline
  \end{tabular*}
 }
 \end{center}
\end{table*}

\begin{figure*}
 \begin{center}
  \includegraphics[width=0.45\textwidth]{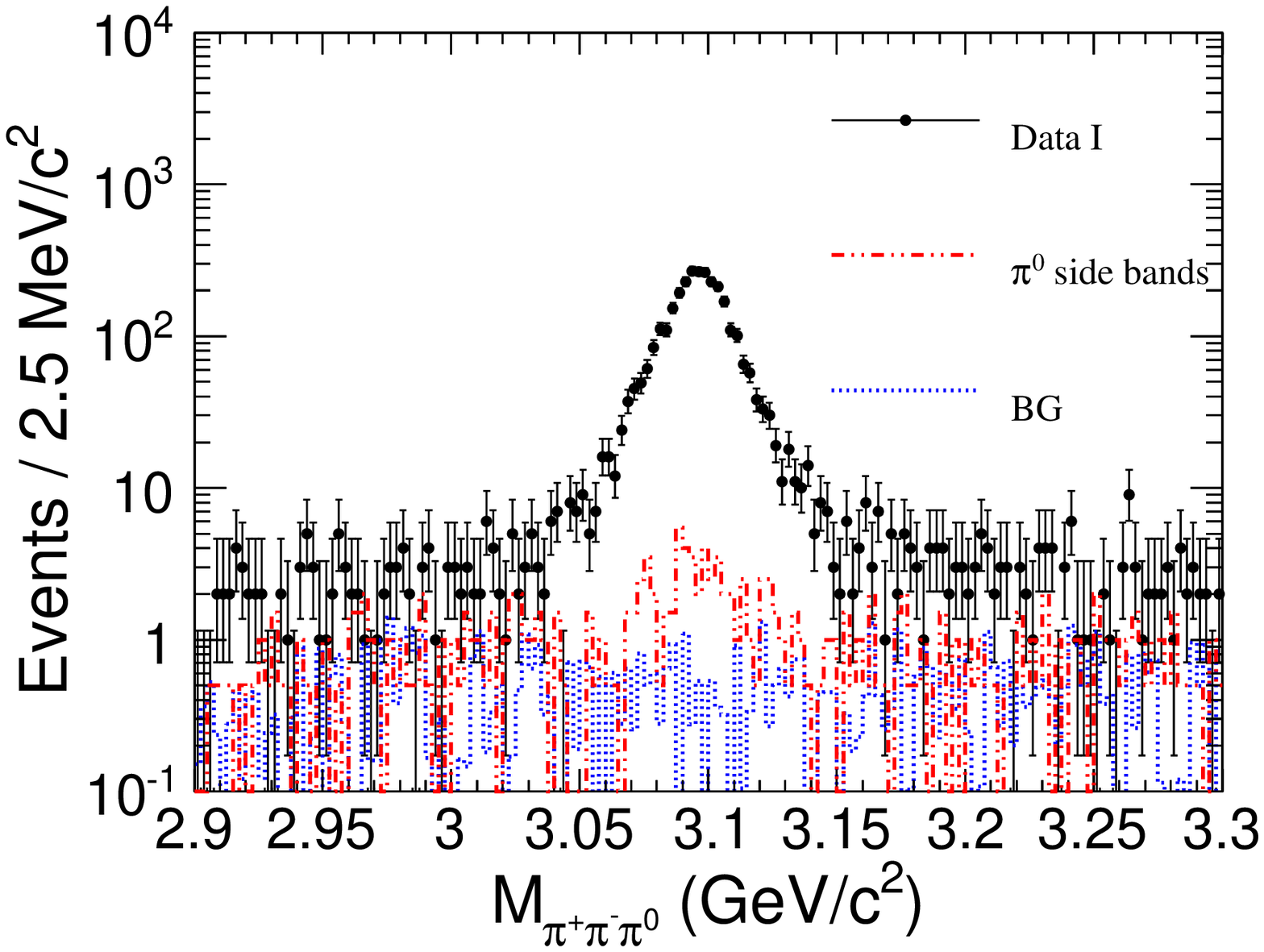}%
  \includegraphics[width=0.45\textwidth]{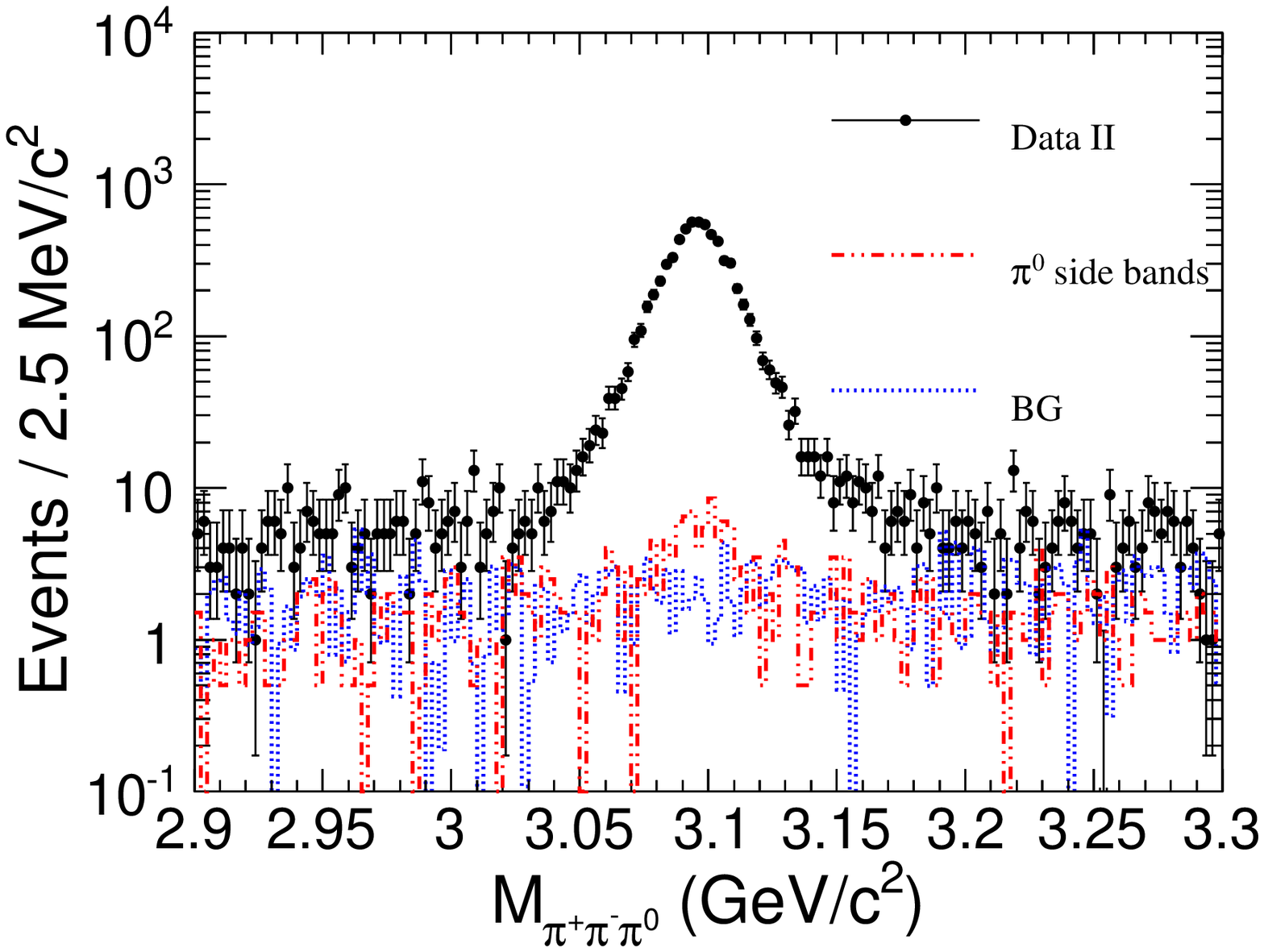}\\%
  \caption{\label{f_m3pi_jpsi}The $\jpsi$ mass spectra in data are shown as black dots with error bars, and the $\pi^0$ side bands and other backgrounds estimated with MC are shown as dashed lines.
}
 \end{center}
\end{figure*}

The result of the untagged ISR method is used to calculate the branching fraction of $\jpsi\to3\pi$. As shown in Fig.~\ref{f_m3pi_jpsi}, the strong $\jpsi$
signals are almost background free. After the $\pi^0$ side bands and other 
backgrounds estimated from Sec.~\ref{s_bg} being subtracted, the number of 
the $\jpsi$ signal in the region (3.05-3.15$\gev$) is obtained by subtracting the $\jpsi$ side bands, (2.9-3.0$\gev$) and (3.2-3.3$\gev$), 
 and it yields $9453\pm102$ signal events for the combined data sets.

To calculate the branching fraction, the total number $N_{\rm tot}$ of $\jpsi$ in data is determined by the 
convolution of the effective luminosity with the cross section of $\ee\to\jpsi$,
\begin{linenomath}
 \begin{equation}
  \sigma_{\ee\to\jpsi}=\frac{12\pi\Gamma_{ee}\cdot \Gamma_{\rm tot}}{(s-M^2)^2+M^2\Gamma_{\rm tot}^2},
 \end{equation}
\end{linenomath}
where $M$ and $\Gamma_{\rm tot}$ are the mass and the full width of $\jpsi$, $\Gamma_{ee} = 
5.53\pm0.10\keV$ is the dileptonic width of $\jpsi$~\cite{c_pdg2014}. The uncertainty of $\Gamma_{ee}$ is the main 
contribution to the total uncertainty of the number of $\jpsi$, and a systematic error of 1.8\% is assigned. Thus the total number of  $\jpsi$ is determined to be $N_{\rm tot} = (2884.7\pm 1.7({\rm stat.}) \pm 60({\rm sys.}))\times10^3$. With the 
above information, the branching fraction of $\jpsi\to3\pi$ is calculated according to:
\begin{linenomath}
 \begin{equation}
  \BR(\jpsi\to3\pi)=\frac{N_{\rm sig}}{N_{\rm tot}\times\epsilon\times\BR(\piz\to\gam\gam)},
 \end{equation}
\end{linenomath}
where $\epsilon=(15.17\pm0.08)\%$ is the selection efficiency, and the branching fraction $\BR(\piz\to\gam\gam)$ is 
taken from PDG~\cite{c_pdg2014}. The result is $\BR(\jpsi\to3\pi)=(2.188\pm0.024({\rm stat.})\pm0.024({\rm 
sys.})\pm0.040({\Gamma_{\rm ee}}))$\%, where the systematic error is taken from Table~\ref{t_sum_sys}. This result is 
consistent with the previous measurements from \babar: ($2.18\pm0.19$)\%~\cite{c_g3pi_babar}, BES: 
($2.10\pm0.12$)\%~\cite{c_3pi_bes}, and BESIII: ($2.137\pm0.063$)\%~\cite{c_3pi_bes3}, with a slightly improved precision.

\section{Cross section results}\label{s_res}
The Born cross section is extracted from the unfolded 3$\pi$ mass spectrum using the relation
\begin{linenomath}
 \begin{equation}
  \label{e_cs}
	\sigma^{\rm Born}(\sqrt{s'})=\frac{(dN/d\sqrt{s'})_{\rm unfolded}}{\varepsilon\cdot dL/d\sqrt{s'}}\cdot\frac{1}{|1+\Pi|^2},
 \end{equation}
\end{linenomath}
where  $1/{|1+\Pi|^2}$ is the vacuum polarization (VP) correction factor,
which cannot be calculated from the first principles. 
Experimental data of $\ee$ annihilating to hadrons is used as input for the 
calculations. Up to $2\GeV$, the VP effects in the propagator of the virtual photon are evaluated with an accuracy 
better than 0.05\%. However, in the vicinity of the narrow  $\omega$ and $\phi$ resonances, the errors are 0.08\% and 
0.3\%~\cite{c_vp_cmd}, respectively.

\begin{figure*}
	\begin{center}
	\includegraphics[width=0.5\textwidth]{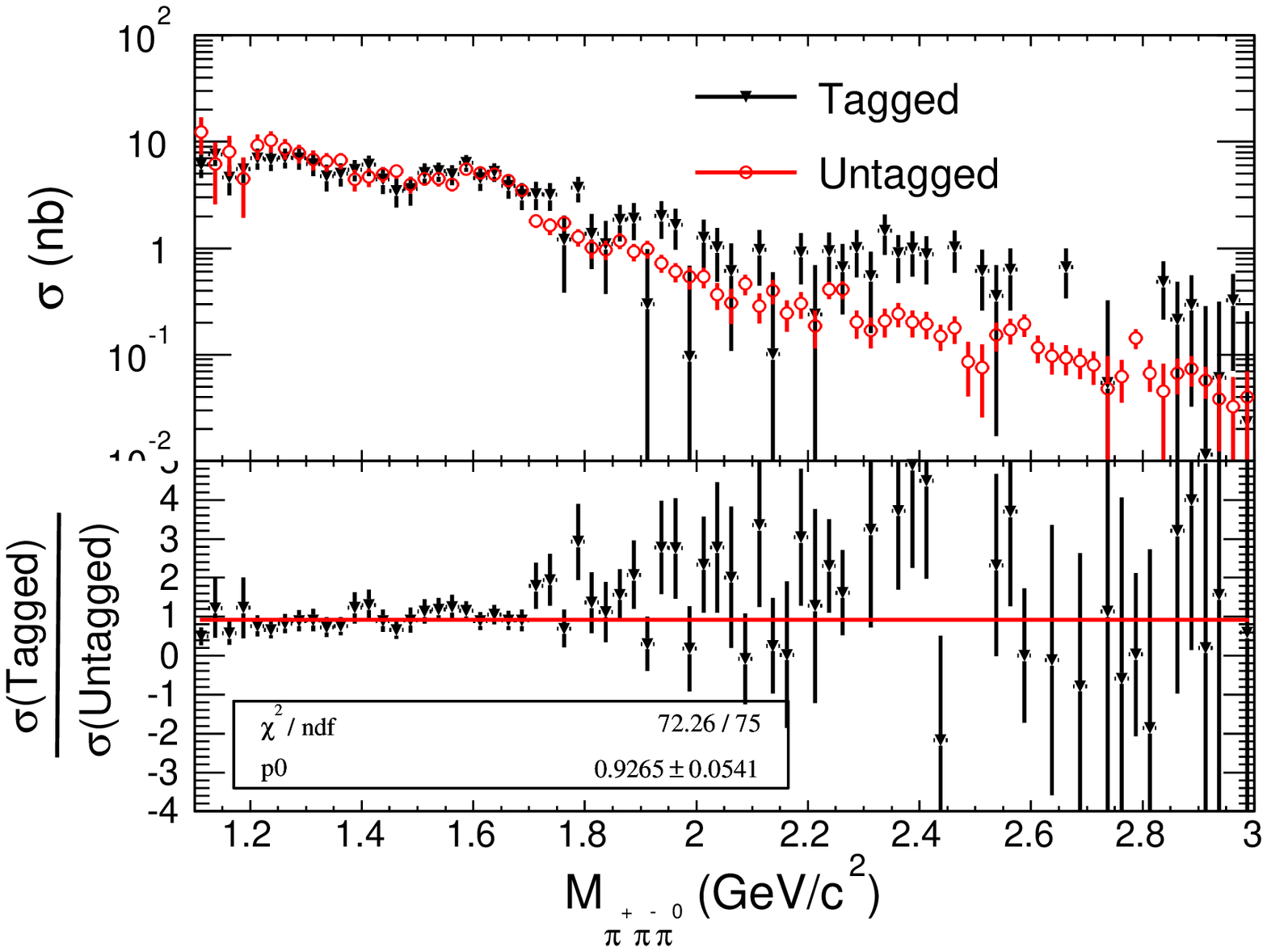}%
	\includegraphics[width=0.5\textwidth]{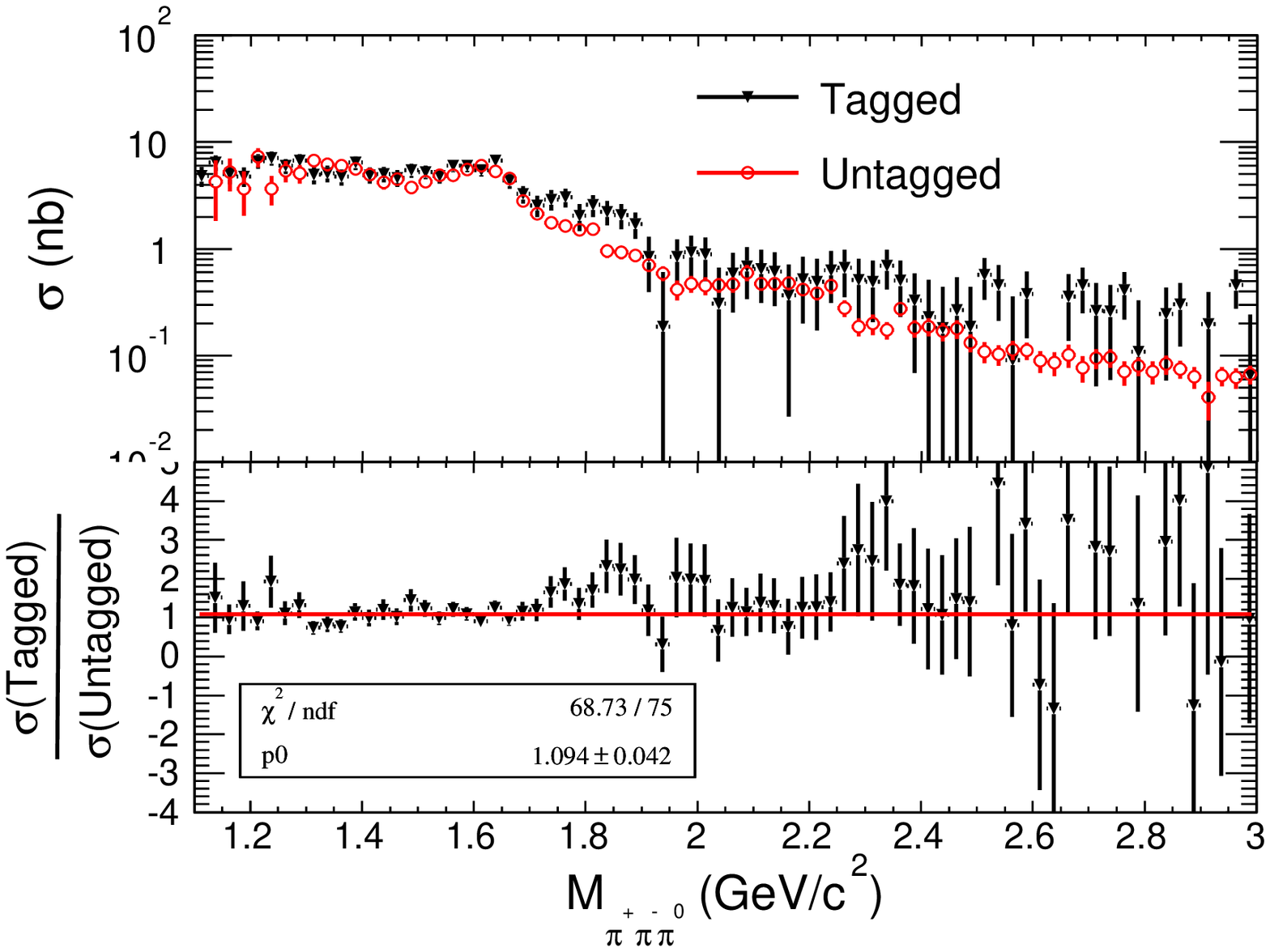}%
	\caption{\label{f_cs_all}Cross sections obtained with the two methods. The left plot is for Data I, and the right is for Data II. The top plots are cross sections for both the tagged and untagged results, while the bottom plots are the ratios of the two.}
	\end{center}
\end{figure*}

\begin{figure*}
 \begin{center}
  \includegraphics[width=0.33\textwidth]{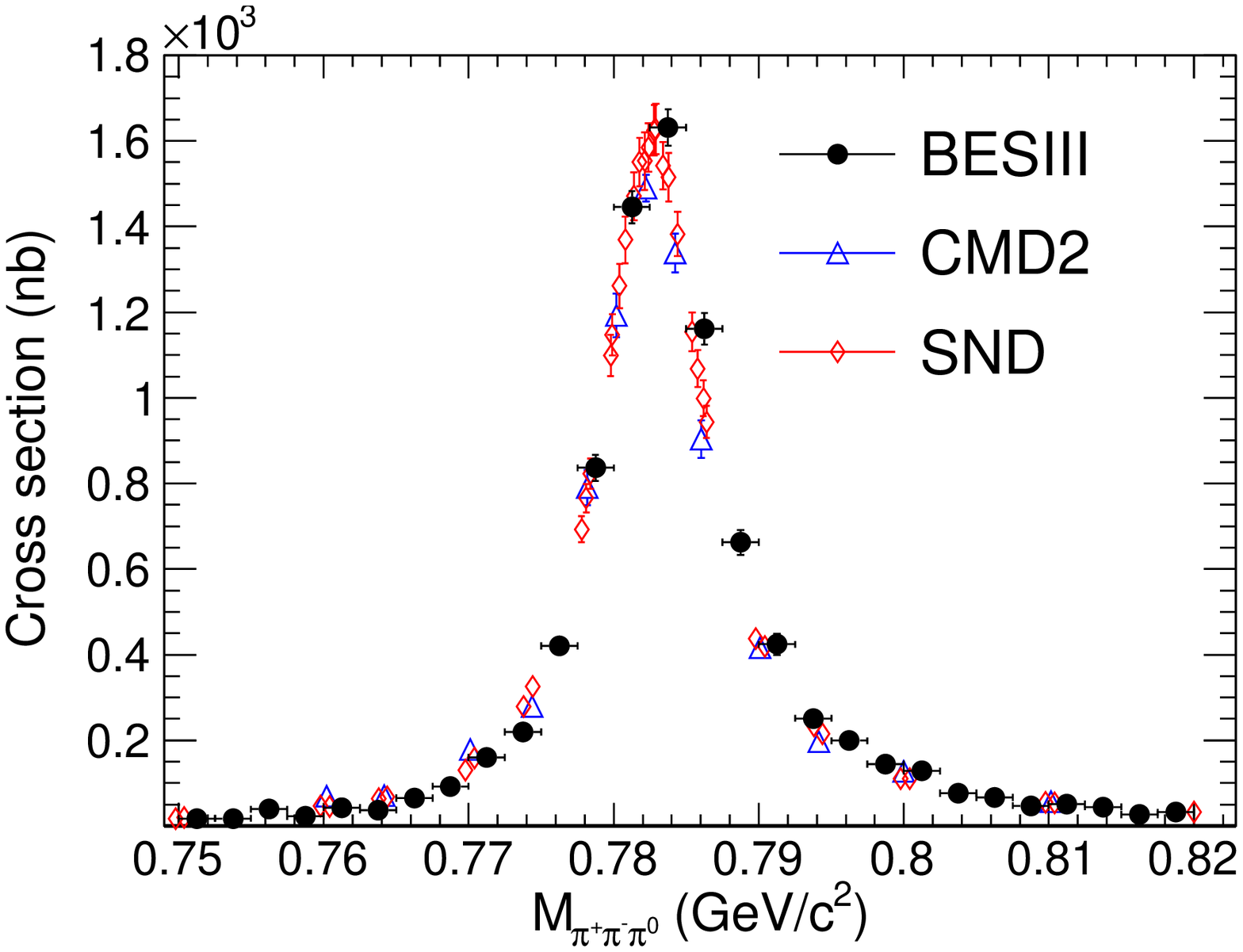}%
  \includegraphics[width=0.33\textwidth]{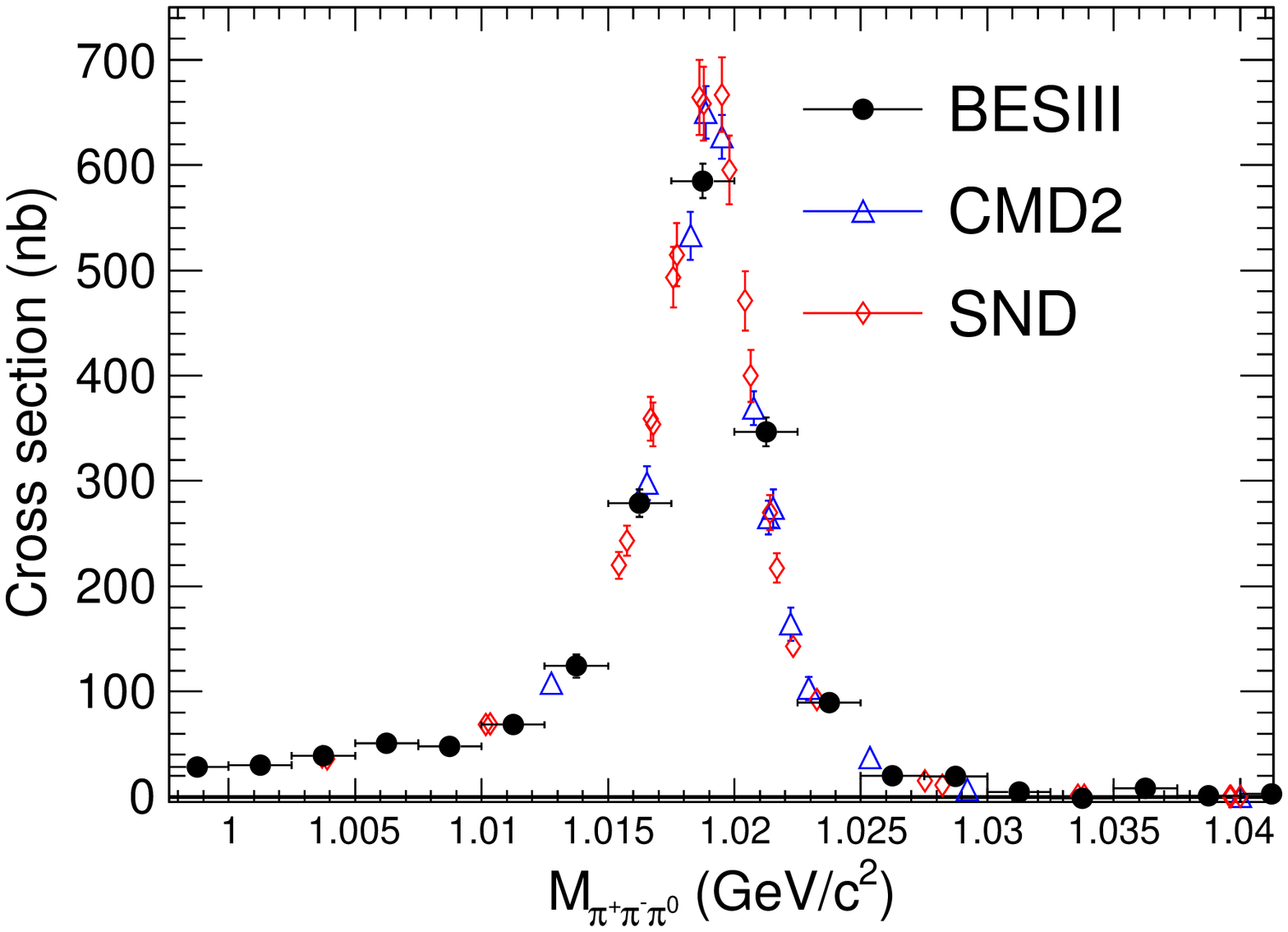}%
  \includegraphics[width=0.33\textwidth]{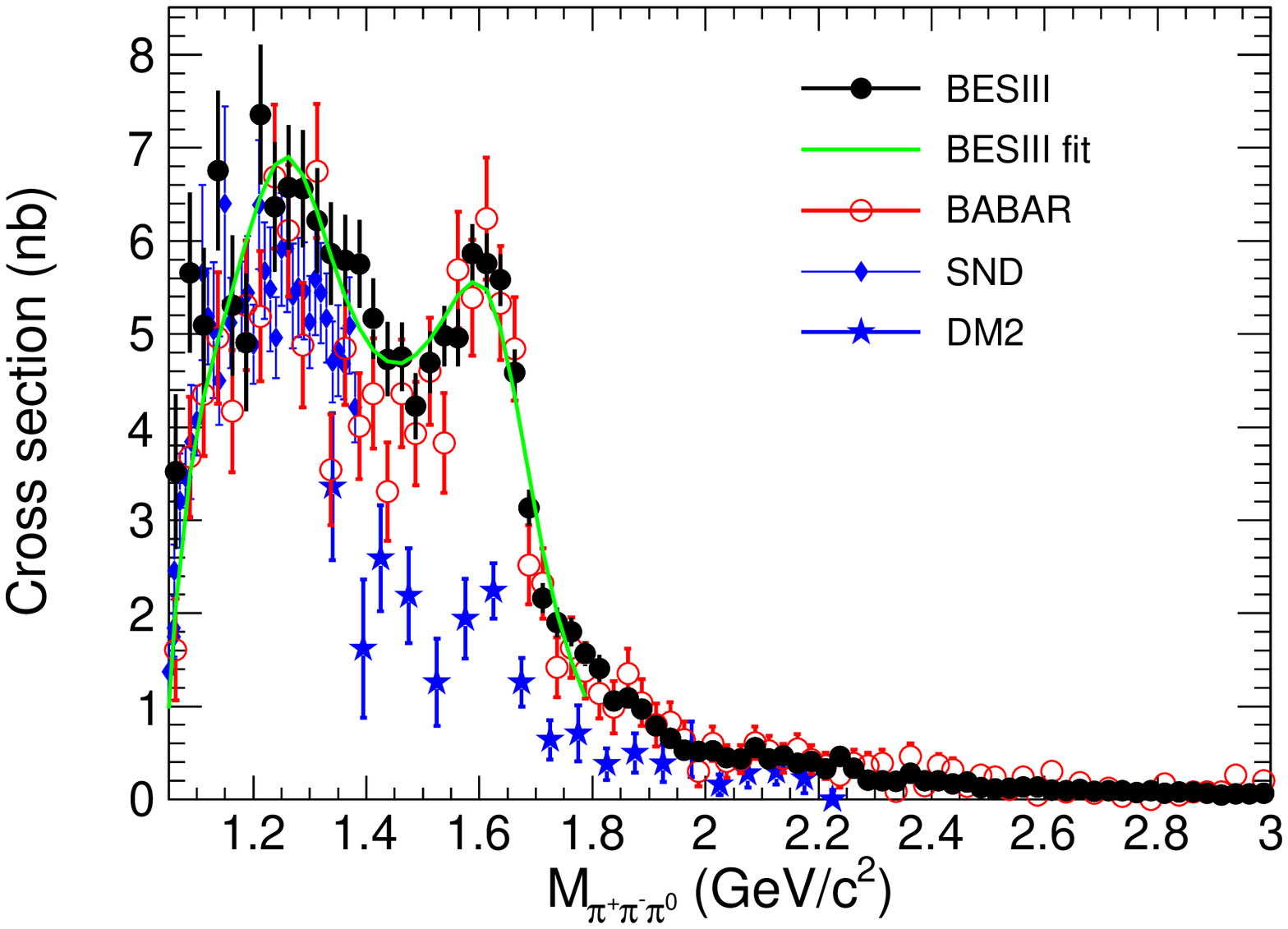}%
 \end{center}
\caption{\label{f_cs_com}Born cross sections and the comparisons with SND, CMD-2, and \babar.
From left to right, they are for the mass regions of $\omega$, $\phi$, and above
$\phi$.}
\end{figure*}

After correcting for the radiator function and the vacuum polarization, the final cross section is obtained. It is 
measured from 0.7 to $3.0\GeV$ using the tagged ISR method, whereas the untagged ISR method is applied from 1.05 to 
$3.0\GeV$. Figure ~\ref{f_cs_all} shows the cross sections as a
function of 3$\pi$ mass.
In the region of overlap (1.05-$3.0\GeV$), the results of the tagged
and the untagged methods are statistically compatible within
$2\sigma$, and are therefore combined according to their statistical errors, as shown in 
Fig.~\ref{f_cs_com} and Table~\ref{t_cs}. The quoted errors correspond to the statistical (including contribution from 
$\piz$ side band subtraction) and the systematic uncertainties. To compare the result with other experiments, the 
dressed cross sections without  VP correction are calculated and shown in Fig.~\ref{f_cs_com}. The results reported here 
agree well with the measurement by \babar, but the values are slightly larger than those of the SND measurement. 
However, they are all consistent within uncertainties. The two resonances $\omegap$ and $\omegapp$, as reported by 
\babar, are clearly confirmed. The so-called $\omegap$ resonance is unusual with respect to the $\omegapp$, since the width is 
expected to increase with a higher excited state.

\begin{table*}\smaller\renewcommand{\arraystretch}{0.99}
 \caption{\label{t_cs}Born cross sections of $\ee\to\pipi\piz$ for the mass region from 0.7 GeV to 3.0 GeV.}
 \begin{center}
	 \begin{tabular}{c|r|c|r|c|r}
   \hline\hline
   $\sqrt{s}$ (GeV)&$\sigma$ (nb)&$\sqrt{s}$ (GeV)&$\sigma$ (nb)&$\sqrt{s}$ (GeV)&$\sigma$ (nb)\\
   \hline
   0.70125&    $1.21\pm2.23\pm0.04$   &  0.88375  &  $13.19\pm4.00\pm0.50$  &  1.21250  &  $7.094\pm0.715\pm0.084$\\
   0.70375&   $-1.74\pm2.54\pm0.13$   &  0.88625  &   $6.61\pm3.40\pm0.30$  &  1.23750  &  $6.143\pm0.664\pm0.102$\\
   0.70625&    $0.92\pm3.03\pm0.03$   &  0.88875  &   $5.98\pm3.29\pm0.14$  &  1.26250  &  $6.345\pm0.634\pm0.113$\\
   0.70875&   $-0.23\pm3.43\pm0.02$   &  0.89125  &  $13.07\pm4.24\pm0.46$  &  1.28750  &  $6.336\pm0.605\pm0.088$\\
   0.71125&    $2.19\pm3.61\pm0.07$   &  0.89375  &  $14.43\pm4.08\pm0.30$  &  1.31250  &  $6.008\pm0.539\pm0.066$\\
   0.71375&   $10.46\pm4.88\pm0.24$   &  0.89625  &  $13.31\pm3.83\pm0.23$  &  1.33750  &  $5.670\pm0.526\pm0.062$\\
   0.71625&    $0.34\pm3.54\pm0.03$   &  0.89875  &  $12.31\pm3.81\pm0.20$  &  1.36250  &  $5.599\pm0.468\pm0.056$\\
   0.71875&   $-0.39\pm4.17\pm0.03$   &  0.90125  &  $10.20\pm3.86\pm0.18$  &  1.38750  &  $5.565\pm0.450\pm0.069$\\
   0.72125&   $11.09\pm5.55\pm0.17$   &  0.90375  &  $12.48\pm3.73\pm0.40$  &  1.41250  &  $4.996\pm0.411\pm0.057$\\
   0.72375&    $1.77\pm5.48\pm0.03$   &  0.90625  &   $9.16\pm3.96\pm0.19$  &  1.43750  &  $4.571\pm0.382\pm0.050$\\
   0.72625&    $9.55\pm4.24\pm0.16$   &  0.90875  &  $12.89\pm3.93\pm0.28$  &  1.46250  &  $4.593\pm0.353\pm0.051$\\
   0.72875&    $9.18\pm5.06\pm0.25$   &  0.91125  &  $11.34\pm4.65\pm0.19$  &  1.48750  &  $4.084\pm0.339\pm0.047$\\
   0.73125&   $12.10\pm4.31\pm0.23$   &  0.91375  &   $7.53\pm3.44\pm0.19$  &  1.51250  &  $4.536\pm0.315\pm0.047$\\
   0.73375&    $3.31\pm5.25\pm0.09$   &  0.91625  &   $8.76\pm3.62\pm0.30$  &  1.53750  &  $4.808\pm0.308\pm0.049$\\
   0.73625&    $2.86\pm3.36\pm0.06$   &  0.91875  &  $11.57\pm3.54\pm0.58$  &  1.56250  &  $4.792\pm0.292\pm0.052$\\
   0.73875&   $11.63\pm6.28\pm0.25$   &  0.92125  &  $13.42\pm3.48\pm0.59$  &  1.58750  &  $5.660\pm0.296\pm0.065$\\
   0.74125&   $16.15\pm5.14\pm0.25$   &  0.92375  &   $4.68\pm4.52\pm0.16$  &  1.61250  &  $5.556\pm0.308\pm0.056$\\
   0.74375&   $13.34\pm5.36\pm0.20$   &  0.92625  &   $9.79\pm3.24\pm0.26$  &  1.63750  &  $5.385\pm0.262\pm0.062$\\
   0.74625&   $15.84\pm5.49\pm0.22$   &  0.92875  &  $10.35\pm3.58\pm0.18$  &  1.66250  &  $4.419\pm0.234\pm0.051$\\
   0.74875&   $15.72\pm5.94\pm0.25$   &  0.93125  &   $8.99\pm3.24\pm0.15$  &  1.68750  &  $3.018\pm0.185\pm0.035$\\
   0.75125&   $17.73\pm6.28\pm0.28$   &  0.93375  &  $10.88\pm3.47\pm0.17$  &  1.71250  &  $2.082\pm0.154\pm0.023$\\
   0.75375&   $16.55\pm6.74\pm0.28$   &  0.93625  &   $7.94\pm3.60\pm0.12$  &  1.73750  &  $1.829\pm0.153\pm0.021$\\
   0.75625&   $38.82\pm8.67\pm0.71$   &  0.93875  &   $4.28\pm3.84\pm0.06$  &  1.76250  &  $1.735\pm0.147\pm0.024$\\
   0.75875&   $23.17\pm7.64\pm0.69$   &  0.94125  &   $9.43\pm3.86\pm0.13$  &  1.78750  &  $1.504\pm0.127\pm0.019$\\
   0.76125&   $40.70\pm9.43\pm1.45$   &  0.94375  &  $11.42\pm3.73\pm0.17$  &  1.81250  &  $1.353\pm0.137\pm0.015$\\
   0.76375&   $38.12\pm9.77\pm1.19$   &  0.94625  &  $14.04\pm3.56\pm0.22$  &  1.83750  &  $1.017\pm0.110\pm0.012$\\
   0.76625&   $62.43\pm12.57\pm1.52$  &  0.94875  &  $17.07\pm3.80\pm0.39$  &  1.86250  &  $1.048\pm0.102\pm0.012$\\
   0.76875&   $91.37\pm13.84\pm1.36$  &  0.95125  &   $7.66\pm2.91\pm0.19$  &  1.88750  &  $0.937\pm0.094\pm0.014$\\
   0.77125&  $157.25\pm18.61\pm3.02$  &  0.95375  &   $7.78\pm3.24\pm0.23$  &  1.91250  &  $0.763\pm0.084\pm0.012$\\
   0.77375&  $218.35\pm22.42\pm3.81$  &  0.95625  &  $10.42\pm4.03\pm0.31$  &  1.93750  &  $0.631\pm0.077\pm0.008$\\
   0.77625&  $415.56\pm21.55\pm7.64$  &  0.95875  &   $8.73\pm3.29\pm0.23$  &  1.96250  &  $0.506\pm0.072\pm0.006$\\
   0.77875&  $828.58\pm27.72\pm12.36$ &  0.96125  &   $7.55\pm3.30\pm0.32$  &  1.98750  &  $0.498\pm0.069\pm0.006$\\
   0.78125& $1405.30\pm31.05\pm19.20$ &  0.96375  &  $11.41\pm3.78\pm0.75$  &  2.01250  &  $0.504\pm0.067\pm0.009$\\
   0.78375& $1534.34\pm33.91\pm21.67$ &  0.96625  &  $19.28\pm4.34\pm0.44$  &  2.03750  &  $0.432\pm0.059\pm0.005$\\
   0.78625& $1084.16\pm30.73\pm14.85$ &  0.96875  &  $7.68\pm4.17\pm0.17$   &  2.06250  &  $0.421\pm0.061\pm0.005$\\
   0.78875&  $618.58\pm25.64\pm8.58$  &  0.97125  &  $18.63\pm4.23\pm0.33$  &  2.08750  &  $0.532\pm0.070\pm0.006$\\
   0.79125&  $399.32\pm22.40\pm5.70$  &  0.97375  &  $18.15\pm4.10\pm0.31$  &  2.11250  &  $0.420\pm0.051\pm0.006$\\
   0.79375&  $239.67\pm18.35\pm4.90$  &  0.97625  &  $19.52\pm4.72\pm0.48$  &  2.13750  &  $0.450\pm0.056\pm0.007$\\
   0.79625&  $184.05\pm16.88\pm5.21$  &  0.97875  &  $11.29\pm3.92\pm0.36$  &  2.16250  &  $0.371\pm0.052\pm0.005$\\
   0.79875&  $136.99\pm13.93\pm2.23$  &  0.98125  &  $12.31\pm3.88\pm0.38$  &  2.18750  &  $0.391\pm0.046\pm0.005$\\
   0.80125&  $122.17\pm13.21\pm1.88$  &  0.98375  &  $11.81\pm3.79\pm0.33$  &  2.21250  &  $0.318\pm0.043\pm0.004$\\
   0.80375&   $72.57\pm10.83\pm1.14$  &  0.98625  &  $15.00\pm3.75\pm0.73$  &  2.23750  &  $0.442\pm0.050\pm0.005$\\
   0.80625&   $62.03\pm8.76\pm1.39$   &  0.98875  &  $15.23\pm4.57\pm0.28$  &  2.26250  &  $0.325\pm0.041\pm0.006$\\
   0.80875&   $45.17\pm7.05\pm1.59$   &  0.99125  &  $15.75\pm4.39\pm0.37$  &  2.28750  &  $0.197\pm0.031\pm0.004$\\
   0.81125&   $49.85\pm8.09\pm1.78$   &  0.99375  &  $26.39\pm4.98\pm0.42$  &  2.31250  &  $0.194\pm0.034\pm0.002$\\
   0.81375&   $41.86\pm7.36\pm1.49$   &  0.99625  &  $24.52\pm5.23\pm0.62$  &  2.33750  &  $0.188\pm0.028\pm0.002$\\
   0.81625&   $25.44\pm5.45\pm0.83$   &  0.99875  &  $28.14\pm5.91\pm0.67$  &  2.36250  &  $0.271\pm0.034\pm0.003$\\
   0.81875&   $29.98\pm6.48\pm0.59$   &  1.00125  &  $29.94\pm6.20\pm0.61$  &  2.38750  &  $0.191\pm0.029\pm0.002$\\
   0.82125&   $31.12\pm6.55\pm0.50$   &  1.00375  &  $38.24\pm7.00\pm0.76$  &  2.41250  &  $0.193\pm0.030\pm0.006$\\
   0.82375&   $27.15\pm6.33\pm0.68$   &  1.00625  &  $50.52\pm7.31\pm1.48$  &  2.43750  &  $0.159\pm0.025\pm0.007$\\
   0.82625&   $23.82\pm6.05\pm0.39$   &  1.00875  &  $49.17\pm8.35\pm1.03$  &  2.46250  &  $0.183\pm0.029\pm0.003$\\
   0.82875&   $29.20\pm6.16\pm0.70$   &  1.01125  &  $69.59\pm9.24\pm2.35$  &  2.48750  &  $0.120\pm0.021\pm0.002$\\
   0.83125&   $10.64\pm4.83\pm0.29$   &  1.01375  & $128.91\pm11.25\pm2.02$ &  2.51250  &  $0.108\pm0.022\pm0.001$\\
   0.83375&   $16.63\pm4.95\pm0.58$   &  1.01625  & $302.66\pm13.57\pm4.12$ &  2.53750  &  $0.115\pm0.020\pm0.001$\\
   0.83625&   $13.95\pm4.20\pm0.42$   &  1.01875  & $591.69\pm14.57\pm8.15$ &  2.56250  &  $0.127\pm0.021\pm0.002$\\
   0.83875&   $24.43\pm5.53\pm0.64$   &  1.02125  & $297.73\pm10.63\pm4.60$ &  2.58750  &  $0.127\pm0.019\pm0.002$\\
   0.84125&   $27.35\pm5.81\pm0.71$   &  1.02375  &  $79.63\pm8.74\pm1.55$  &  2.61250  &  $0.094\pm0.017\pm0.001$\\
   0.84375&   $10.97\pm4.17\pm0.19$   &  1.02625  &  $18.47\pm7.43\pm0.54$  &  2.63750  &  $0.088\pm0.017\pm0.001$\\
   0.84625&   $20.91\pm4.64\pm0.35$   &  1.02875  &  $16.82\pm6.38\pm0.63$  &  2.66250  &  $0.102\pm0.019\pm0.001$\\
   0.84875&   $14.13\pm4.53\pm0.47$   &  1.03125  &   $4.53\pm5.71\pm0.37$  &  2.68750  &  $0.083\pm0.017\pm0.001$\\
   0.85125&   $15.34\pm3.69\pm0.46$   &  1.03375  &  $-1.25\pm3.81\pm0.06$  &  2.71250  &  $0.089\pm0.016\pm0.001$\\
   0.85375&   $12.77\pm4.08\pm0.19$   &  1.03625  &   $7.48\pm4.27\pm0.29$  &  2.73750  &  $0.091\pm0.017\pm0.001$\\
   0.85625&   $10.81\pm4.38\pm0.59$   &  1.03875  &   $1.23\pm3.28\pm0.08$  &  2.76250  &  $0.070\pm0.015\pm0.001$\\
   0.85875&   $12.94\pm4.96\pm0.26$   &  1.04125  &   $2.68\pm4.21\pm0.09$  &  2.78750  &  $0.092\pm0.014\pm0.001$\\
   0.86125&   $20.54\pm4.17\pm0.65$   &  1.04375  &   $0.08\pm2.46\pm0.08$  &  2.81250  &  $0.067\pm0.014\pm0.001$\\
   0.86375&   $10.17\pm4.31\pm0.24$   &  1.04625  &  $-0.90\pm3.08\pm0.17$  &  2.83750  &  $0.079\pm0.015\pm0.001$\\
   0.86625&   $11.80\pm3.83\pm0.28$   &  1.04875  &   $3.64\pm2.87\pm0.19$  &  2.86250  &  $0.074\pm0.013\pm0.001$\\
   0.86875&    $3.89\pm3.85\pm0.11$   &  1.06250  &   $3.351\pm0.788\pm0.070$  &  2.88750  &  $0.066\pm0.012\pm0.001$\\
   0.87125&   $18.25\pm4.46\pm0.65$   &  1.08750  &   $5.417\pm0.815\pm0.116$  &  2.91250  &  $0.048\pm0.012\pm0.002$\\
   0.87375&    $4.47\pm3.37\pm0.17$   &  1.11250  &   $4.889\pm0.790\pm0.071$  &  2.93750  &  $0.059\pm0.012\pm0.001$\\
   0.87625&    $6.48\pm3.56\pm0.20$   &  1.13750  &   $6.494\pm0.810\pm0.156$  &  2.96250  &  $0.059\pm0.012\pm0.001$\\
   0.87875&    $8.32\pm3.64\pm0.26$   &  1.16250  &   $5.106\pm0.722\pm0.088$  &  2.98750  &  $0.062\pm0.012\pm0.002$\\
   0.88125&    $7.30\pm3.80\pm0.30$   &  1.18750  &   $4.730\pm0.698\pm0.136$  &           &\\ 
   \hline \hline
  \end{tabular}
 \end{center}
\end{table*}

In order to calculate the $\pipi\piz$ contribution to $a_\mu$, the cross section including FSR, $\ee\to\pipi\piz\gam_{\rm FSR}$, is obtained by the formula
\begin{equation}
	\sigma^{\rm FSR}=\sigma^{\rm Born}\left(1+F(s)\frac{\alpha}{\pi}\right),
\end{equation}
where $\alpha$ is the fine-structure constant, and $F(s)$ is the FSR correction factor from Ref.~\cite{c_FSRamu}.
Using the cross section $\ee\to\pipi\piz\gam_{\rm FSR}$ as input for Eq.~(\ref{eq_dis}), the contribution to $a_\mu$ from the $\pipi\piz$ process is calculated to be $a_\mu^{3\pi}(0.7-3.0 \,{\rm GeV})=(49.77\pm0.53\pm0.17)\times10^{-10}$. This result is larger than previous calculations~\cite{c_thomas_2018} which are using both measurement and fits. However, the difference is less than 2$\sigma$.

\section{Summary}\label{s_sum}
Based on $2.93\ifb$ of $\ee$ collision data taken at $\sqrt{s}=3.773\GeV$ with the BESIII detector, $\ee\to\pipi\piz$ is 
studied using the ISR method. The cross section of $\ee\to3\pi$ is obtained for $\sqrt{s}$ from 0.7 to $3.0\GeV$, and 
$a_\mu^{3\pi}$ is calculated to be $(49.77\pm0.53\pm0.17)\times10^{-10}$. 
This is the first calculation using data from a single experiment, and 
the uncertainty of $a_\mu^{3\pi}$ is reduced by 40\%.
In this energy region, a clear indication of 
the $\omegapp$ around $1.6\gev$ is found.  
This  agrees with \babar~\cite{c_g3pi_babar} but is significantly higher than the DM2~\cite{c_3pi_dm2} result. 
The 3$\pi$ mass spectrum below $1.8\gev$ can be well described by the coherent sum of the resonances $\omega, 
\phi$, $\omegap$, and $\omegapp$. Product branching fractions are extracted from the fit
and are consistent 
with both the PDG~\cite{c_pdg2014} values and the \babar~\cite{c_g3pi_babar} result. A study of $\jpsi\to3\pi$ is 
performed with untagged data above $3.0\gev$. The branching fraction is calculated, showing good agreement with 
previous results~\cite{c_g3pi_babar,c_3pi_bes,c_3pi_bes3} with slightly improved precision. 

\section*{Acknowledgement}
The BESIII collaboration thanks the staff of BEPCII and the IHEP computing center for their strong support. This work is supported in part by National Key Basic Research Program of China under Contract No. 2015CB856700; National Natural Science Foundation of China (NSFC) under Contracts Nos. 11625523, 11635010, 11735014, 11822506, 11835012; the Chinese Academy of Sciences (CAS) Large-Scale Scientific Facility Program; Joint Large-Scale Scientific Facility Funds of the NSFC and CAS under Contracts Nos. U1532257, U1532258, U1732263, U1832207; CAS Key Research Program of Frontier Sciences under Contracts Nos. QYZDJ-SSW-SLH003, QYZDJ-SSW-SLH040; 100 Talents Program of CAS; INPAC and Shanghai Key Laboratory for Particle Physics and Cosmology; ERC under Contract No. 758462; German Research Foundation DFG under Contracts Nos. Collaborative Research Center CRC 1044, FOR 2359; Istituto Nazionale di Fisica Nucleare, Italy; Koninklijke Nederlandse Akademie van Wetenschappen (KNAW) under Contract No. 530-4CDP03; Ministry of Development of Turkey under Contract No. DPT2006K-120470; National Science and Technology fund; STFC (United Kingdom); The Knut and Alice Wallenberg Foundation (Sweden) under Contract No. 2016.0157; The Royal Society, UK under Contracts Nos. DH140054, DH160214; The Swedish Research Council; U. S. Department of Energy under Contracts Nos. DE-FG02-05ER41374, DE-SC-0010118, DE-SC-0012069; University of Groningen (RuG) and the Helmholtzzentrum fuer Schwerionenforschung GmbH (GSI), Darmstadt.

\end{document}